\documentclass[journal=psy]{CUP-JNL-DTM}%

\usepackage{graphicx}
\usepackage{multicol,multirow}
\usepackage{amsmath,amssymb,amsfonts}
\usepackage{mathrsfs}
\usepackage{amsthm}
\usepackage{rotating}
\usepackage{appendix}
\usepackage{ifpdf}
\usepackage[T1]{fontenc}
\usepackage{newtxtext}
\usepackage{newtxmath}
\usepackage{textcomp}
\usepackage{xcolor}
\usepackage{lastpage}
\usepackage[colorlinks,allcolors=blue]{hyperref}

\theoremstyle{definition}

\numberwithin{equation}{section}

\usepackage{amsmath}
\usepackage[nopatch]{microtype}
\usepackage[utf8]{inputenc}
\usepackage{newunicodechar}
\newunicodechar{−}{-}

\usepackage{amsfonts}       
\usepackage{nicefrac}       
\usepackage{microtype}      
\usepackage{lipsum}
\usepackage{comment}

\usepackage[english]{babel}
\usepackage{multirow}
\usepackage{amsthm}
\usepackage{amssymb}
\usepackage{bbm}
\usepackage{float}
\usepackage{subcaption}
\usepackage{multirow}
\usepackage{makecell}
\usepackage{mathtools}
\usepackage{algorithmic}
\usepackage{bm}
\usepackage{physics}
\usepackage{appendix}
\usepackage{enumitem}
\usepackage{csquotes}
\usepackage{cancel}
\usepackage{setspace}
\allowdisplaybreaks

\newcommand{\indFct}[1]{\ensuremath{\mathbbm{I}\left(#1\right)}}

\renewcommand{\real}{\ensuremath{\mathbbm{R}}}
\renewcommand{\v}[1]{\ensuremath{\bm{#1}}}

\newcommand{\bfmu}{\boldsymbol{\mu}}
\newcommand{\bftau}{\boldsymbol{\tau}}
\newcommand{\bftheta}{\boldsymbol{\theta}}
\newcommand{\bfSigma}{\boldsymbol{\Sigma}}
\newcommand{\bfOmega}{\boldsymbol{\Omega}}
\newcommand{\bfXi}{\boldsymbol{\Xi}}
\newcommand{\bfalpha}{\boldsymbol{\alpha}}
\newcommand{\bfdelta}{\boldsymbol{\delta}}

\newcommand{\bfvarphi}{\boldsymbol{\varphi}}
\newcommand{\bfeta}{\boldsymbol{\eta}}
\newcommand{\bfbeta}{\boldsymbol{\beta}}
\newcommand{\bfupsilon}{\boldsymbol{\upsilon}}
\newcommand{\bfy}{\boldsymbol{y}}
\newcommand{\bfz}{\boldsymbol{z}}
\newcommand{\bfY}{\boldsymbol{Y}}
\newcommand{\bfD}{\boldsymbol{D}}
\DeclareMathOperator{\argmax}{argmax}
\DeclareMathOperator{\Bdiag}{Bdiag}

\newcommand{\editMark}[1]{#1}

\jname{Psychometrika}
\articletype{ARTICLE TYPE}
\jyear{YEAR}

\begin{document}

\begin{Frontmatter}

\title[Ordinal Unfolding Models]{Modeling Ordinal Survey Data with Unfolding Models}

\author[1]{Rayleigh Lei}
\author[2]{Abel Rodriguez}

\address[1]{\orgdiv{Center for Statistical Training and Consulting}, \orgname{Michigan State University}, \orgaddress{\city{East Lansing}, \postcode{48824}, \state{Michigan},  \country{United States of America}}
\email{rayleigh@umich.com}}

\address[2]{\orgdiv{Department of Statistics}, \orgname{University of Washington}, \orgaddress{\city{Seattle}, \postcode{98195}, \state{Washington},  \country{United States of America}}. \email{abelrod@uw.com}}

\authormark{Lei and Rodriguez}

\abstract{Surveys that rely on ordinal polychotomous (Likert-like) items are widely employed to capture individual preferences because they allow respondents to express both the direction and strength of their preferences. Latent factor models traditionally used in this context implicitly assume that the response functions (the cumulative distribution of the ordinal outcome) are monotonic on the latent trait. This assumption can be too restrictive in several application areas, including in political science and marketing.  In this work, we propose a novel ordinal probit unfolding model that can accommodate both monotonic and non-monotonic response functions. The performance of the model is illustrated through simulations and through the analysis of an immigration attitude survey conducted in the United States.}

\end{Frontmatter}

\section{Introduction}

The recovery of latent traits (such as attitudes, perceptions, opinions or preferences) from polychotomous ordinal (Likert-like) responses is a common task in the social sciences and beyond, e.g., see \citet{joreskog2001factor} and \citet{bartholomew2011latent}.  So-called graded response models (GRMs,  \citealp{SamejimaEstimationLatentAbility1968}), which are extensions of latent factor models that rely on cumulative-link multinomial regression (e.g., see \citealp{bartholomew2011latent} and \citealp{AgrestiAnalysisOrdinalCategorical2010}), are a widely used tool in this context because of their interpretability and relative ease of computation.  GRMs can also be seen as generalizations of dichotomous item response theory models (IRTs, e.g., see \citealp{fox2010bayesian}).  Additional early references include \citet{bartholomew1980factor} and \citet{muthen1983latent}.  Examples of software packages that implement GRMs include \citet{RizopoulosLtmPackageLatent2006}, \citet{CurtisBUGSCodeItem2010}, \citet{LiuAppliedOrdinalLogistic2016},  \citet{BurknerBrmsPackageBayesian2017} and \citet{VincentEtAlPymcdevsPymcexamplesDecember2022}.

When the latent trait is assumed to be univariate, a well-known shortcoming of GRMs is the implicit assumption that the response functions, which describe the  cumulative distribution of the ordinal response as a function of the  the latent trait, are monotonic.  While this assumption is sensible in some settings (e.g., in educational testing, where $\beta$ represents the ability of an individual on a particular subject and $y$ the level of correctness of a response to an item in a test), it might be too restrictive in others (e.g., in political science or marketing applications, see \citealp{duck2022ends} and \citealp{LeiRodriguezNovelClassUnfolding2023}). The monotonicity of the response functions is a byproduct of the underlying geometry implicit in the definition of GRMs. Extending the model to incorporate multivariate latent spaces often fails to address the underlying lack of model fit while complicating interpretation (e.g., see \citealp{yu2021spatial} for an illustration).

An alternative to GRMs is the  Generalized Graded Unfolding Model (GGUM) introduced in \citet{roberts2000general}.  The key insight behind the construction of the GGUM is that, in the context of one-dimensional latent spaces, individuals might disagree with a particular statement when their own preferences lie too far in either direction from a reference or ``ideal'' point.  \citet{roberts2000general} describe this phenomenon as individuals potentially disagreeing either ``from above'' or ``from below'', and proceed to construct their model for the observed preferences by ``folding'' the outcome space of a model with monotonic response functions so that extreme  categories on either end of the subjective (latent) scale map to a single observed category.

Arguably, GGUMs have become the most popular ideal point model for Likert-like items in the psychometric literature.  Accordingly, various implementations, both frequentist and Bayesian, have been developed (e.g., see \citealp{roberts2006ggum2004}, \citealp{DeLaTorreStarkChernyshenko2006},\citealp{chalmers2012mirt}, \citealp{KingRoberts2015}, \citealp{WangDeLaTorreDrasgow2015}, \citealp{TuZhangAngraveSun2021} and \citealp{duck2022ends}).  Nonetheless, computation for GGUMs remains challenging, particularly in Bayesian settings.  Furthermore, recent evidence suggests that, while GGUMs allows for non-monotonic response functions, the shape of their response functions might be too restrictive (e.g., see \citealp{LeiRodriguezNovelClassUnfolding2023,lei2025logit}, as well as Sections \ref{sec:GRMGGUMreview} and \ref{sec:results}).

In this paper, we introduce a novel unfolding model for polychotomous ordinal data that builds on the random utility framework of \citet{McFaddenModelingChoiceResidential1978}, and discuss how to perform statistical inference under a Bayesian framework. \editMark{As is the case for the previous models discussed, our model assumes that the latent trait is univariate.}  The resulting ordinal probit unfolding model (OPUM) is a \editMark{spatial model \citep{EnelowHinichSpatialTheoryVoting1984}} that extends the approach introduced in \citet{LeiRodriguezNovelClassUnfolding2023} for the analysis of roll-call votes in the U.S.\ Congress.  We demonstrate that OPUM's are simpler to interpret and can provide better complexity-adjusted fit than GGUM's in real datasets.  We also demonstrate that OPUMs have computational advantages over GGUM's because they allow for Markov chain Monte Carlo algorithms that require \editMark{minimal tuning} and are, therefore, easier to employ by practitioners without deep knowledge of computational methods.  \editMark{For similar reasons, OPUM is easier to extend to complex hierarchical setting (for an example, see \citealp{LeiRodriguezNovelClassUnfolding2023}, who develop a dynamic version of the model in the dichotomous case).}

The remainder of the paper is organized as follows. Section \ref{sec:GRMGGUMreview} introduces our notation and provides a more detailed review of various methodologies available in the literature. Section \ref{sec:model} introduces OPUMs and discusses their properties. Section \ref{sec:sampling} describes a Markov chain Monte Carlo algorithm to sample from the posterior distribution of our model. Section \ref{sec:results} presents an application of the model to an inmigration attitude survey that illustrates the advantages of the OPUMs over existing alternatives.  We conclude in Section \ref{sec:discussion} by with a brief discussion that includes future research directions.

\section{A brief review of GRMs and GGUMs}\label{sec:GRMGGUMreview}

Formally, our interest is on modeling an $I \times J$ matrix $\bfY$ of observed values such that $y_{i,j} \in \{ 0, 1, \ldots, K_j \}$ represents the ordinal response of subject $i$ to item $j$.  Our main inferential goal is to learn a latent trait $\beta_i \in \real$ that captures the dependence across the columns of the matrix $\bfY$.  For example, in the application that we discuss in Section \ref{sec:results}, one such item corresponds to the prompt, ``All undocumented immigrants currently living in the U.S. should be required to return to their home country", with possible responses being ``Strongly Disagree" ($y_{i,j}=0$), ``Disagree" ($y_{i,j}=1$), ``Neither" ($y_{i,j}=2$), ``Agree" ($y_{i,j}=3$), and ``Strongly Agree" ($y_{i,j}=4$). 
A GRM for this data is defined through a collection of $K_j$ dichotomous item response models of the form
\begin{align}\label{eq:GRM0}
    \Pr(y_{i,j} \le k \mid \beta_i) &= G\left(\alpha_j \left\{\beta_i - \mu_{j,k}\right\} \right) ,  &  k &=0,\ldots, K_j-1 ,
\end{align}
where $G(\cdot)$ is an appropriate link function, and $\alpha_j$ and $\mu_{j,0} < \ldots < \mu_{j,K_j-1}$ are parameters specific to item $j$. Then, the likelihood function is given by 
\begin{align*}
    f\left(\bfY \mid \{\beta_i\}, \{ \mu_{j,k} \}, \{ \alpha_j \}\right) &= \prod_{i=1}^{I}\prod_{j=1}^{J} \prod_{k=0}^{K_j} \left\{ \theta_{i,j}^{GRM}(k) \right\}^{\indFct{y_{i,j}=k}} ,
\end{align*}
where 
\begin{align}\label{eq:GRM}
    \theta^{GRM}_{i,j}(k) = \Pr(y_{i,j} = k \mid \beta_i) = \begin{cases}
     G\left(\alpha_j \left\{\beta_i - \mu_{j,0} \right\} \right) & k = 0 , \\
     G\left(\alpha_j \left\{\beta_i - \mu_{j,k} \right\} \right) - G\left(\alpha_j \left\{\beta_i - \mu_{j,k - 1} \right\} \right)  & 1 \le k \le K_j-1 , \\
    1 - G\left(\alpha_j \left\{\beta_i - \mu_{j,K_j-1} \right\} \right)  & k = K_j ,
    \end{cases}
\end{align}
and $\indFct{\cdot}$ stands for the indicator function. Note that $\Pr(y_{i,j} \le k \mid \beta_i)$ is guaranteed to be monotonic as a function of $\beta_i$ because the link function $G$ must be monotonic for the factor model in \eqref{eq:GRM0} to be well defined.  A widely used version of this model is the logistic GRM, where $G(t) = \left(1 + \exp\{ -t \} \right)^{-1}$. 

It is well known that GRMs can be motivated through a latent variable representation \parencite{anderson1981regression,AgrestiAnalysisOrdinalCategorical2010}.  In particular, let $z_{i,j} = -\alpha_j \beta_i  + \epsilon_{i,j}$, where the $\epsilon_{i,j}$s are independent and identically distributed according to the distribution function $G$, and define $y_{i,j} \le k$ if and only if $ z_{i,j} \le \tau_{j,k}$ for a sequence of thresholds $\tau_{j,0} < \tau_{j,1} < \cdots < \tau_{J,K_j-1} < \tau_{J,K_j} = \infty$.  Then, integrating over $z_{i,j}$, we find that $\Pr(y_{i,j} = k \mid \beta_i)$ takes the form \eqref{eq:GRM} with $\mu_{j,k} = -\tau_{j,k}/\alpha_j$ for $k < K_j$.  This latent representation is key to many computational implementations of GRMs and facilitates the interpretation of the model.  In particular, this representation suggests that $\mu_{j,0}, \ldots, \mu_{j,K_j-1}$ are akin to difficulty parameters in dichotomous models and $\alpha_j$ is akin to a discrimination parameters.

In contrast, a GGUM for the same set of observations is  defined directly on $ \theta^{GGUM}_{i,j}(k) = \Pr(y_{i,j} = k \mid \beta_i)$, which is given by
\begin{multline}\label{eq:GGUM}
    \Pr(y_{i,j} = k \mid \beta_i) = \\
    \frac{\exp\left\{ \alpha_j \left[  k (\beta_i - \mu_j) - \sum_{l=0}^{k} \tau_{j,l} \right] \right\} + \exp\left\{ \alpha_j \left[  \left(2K_j+1 - k\right) (\beta_i - \mu_j) - \sum_{l=0}^{2K_j+1 - k} \tau_{j,l} \right] \right\}}
    {\sum_{k'=0}^{K_j} \exp\left\{ \alpha_j \left[  k' (\beta_i - \mu_j) - \sum_{l=0}^{k'} \tau_{j,l} \right] \right\} + \exp\left\{ \alpha_j \left[  \left(2K_j+1 - k'\right) (\beta_i - \mu_j) - \sum_{l=0}^{2K_j+1 - k'} \tau_{j,l} \right] \right\}} .
\end{multline}
In this expression, $y_{i,j} = 0$ corresponds to the strongest level of disagreement (and, conversely, $y_{i,j} = K_j$ corresponds to the strongest level of agreement), and the parameters $\tau_{j,0}, \ldots, \tau_{j,K_j+1}$ satisfy $\tau_{j, 0} = \tau_{j,K_j + 1} = 0$ and $\tau_{j,k} = - \tau_{j,2K_j+2-k }$ for $k \neq 0$.  
\begin{figure}
    \centering
    \includegraphics[width=0.5\linewidth]{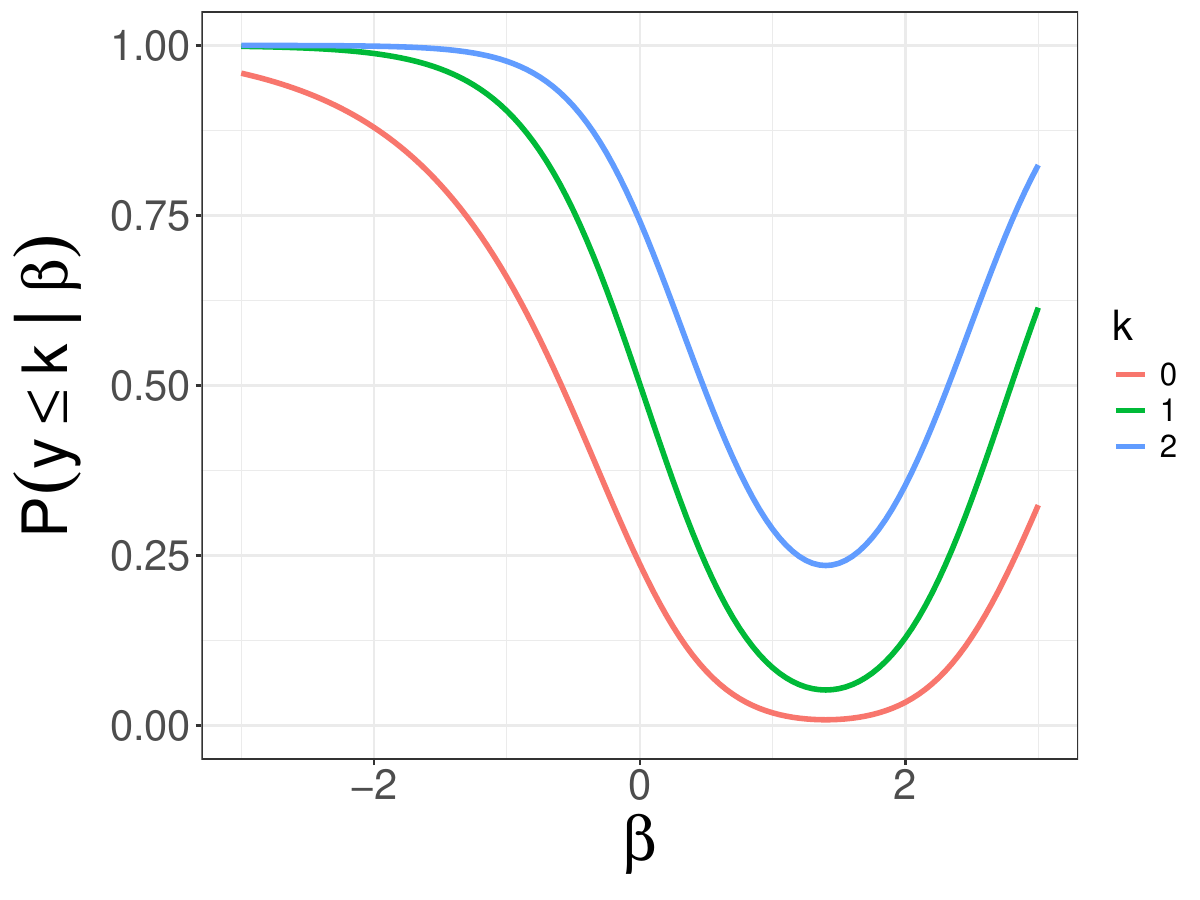}
    \caption{Response function for a BGGUM where $K_j=3$, $\alpha_j = 1$, $\mu_j = 1.1$, $\v{\tau} = (0, -1.2, -1.0, -0.7)$.}
    \label{fig:bggum_cdf}
\end{figure}

\begin{figure}[!t]
    \centering
    \includegraphics[width = 0.5\textwidth]{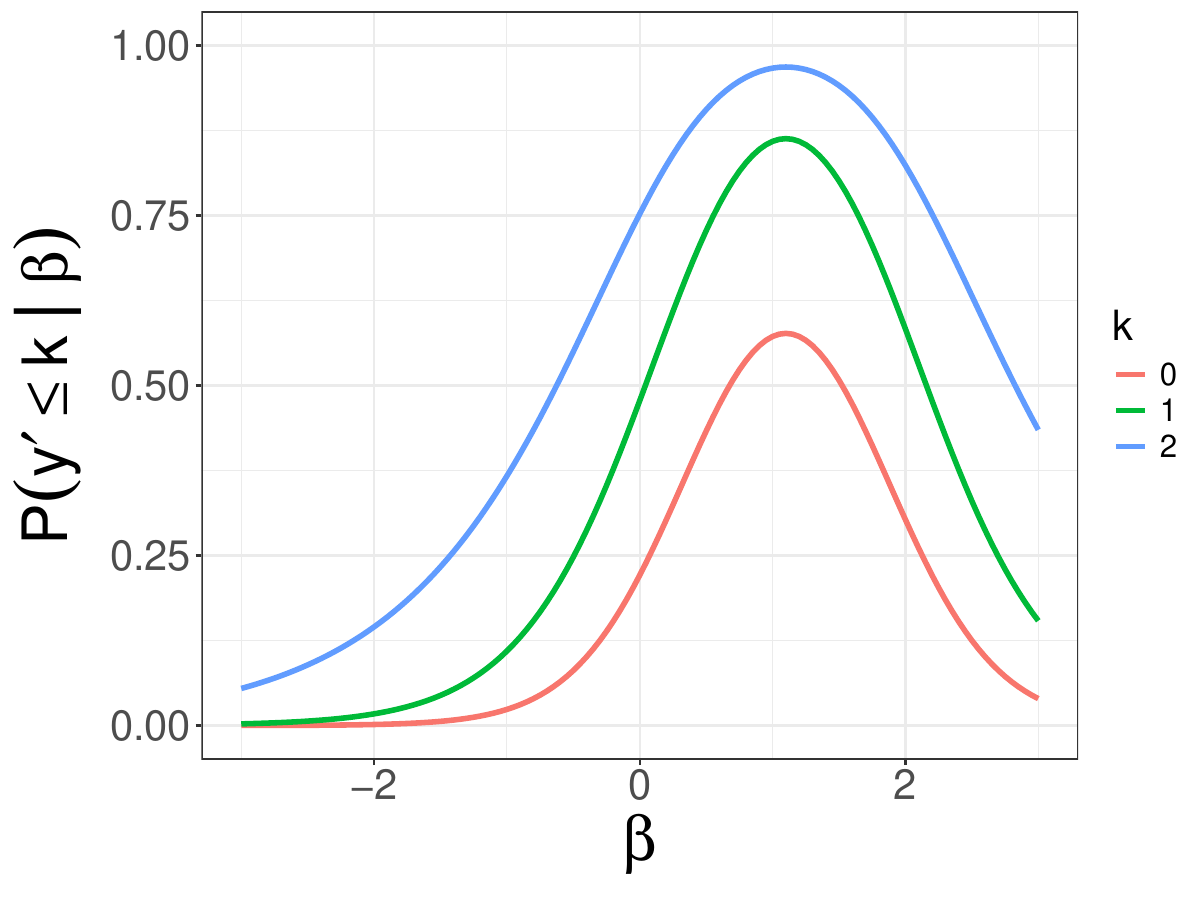}
    \caption{\editMark{Implied response functions for the negated version of the item  in Figure \ref{fig:bggum_cdf}, under the assumption that respondents answer consistently to both versions.}}\label{fig:flipped_scale}
\end{figure}

Unlike GRM's, $\Pr(y_{i,j} \le k \mid \beta_i)$ is not necessarily monotonic on $\beta_i$ for GGUM's (see Figure \ref{fig:bggum_cdf}).  Furthermore, note that the response functions associated with GGUM's always have a minimum at $\beta_i = \mu_j$, are strictly decreasing in the interval $(-\infty,\mu_j)$, strictly increasing in $(\mu_j,\infty)$, and that 
$$
\lim_{\beta_i \to \infty} \Pr(y_{i,j}=0 \mid \beta_i) = \lim_{\beta_i \to -\infty} \Pr(y_{i,j}=0 \mid \beta_i) = 1.
$$
(See our supplementary materials). We argue that this last feature is a key shortcoming of GGUM's, as it implies that they cannot accurately model response functions of items for which individuals on both extremes of the latent scale \textit{strongly agree} with the item.  This makes the inferences generated by the model heavily dependent on how an item has been formulated. \editMark{For a concrete example related to the dataset analyzed in Section \ref{sec:realdata}, consider two versions of the same item: a ``positively'' worded version, ``There are more important issues to address than undocumented immigration'', and a ``negatively'' worded one such as ``Undocumented immigration is an important issue that must be addressed''. Barring inconsistency issues (e.g., see \citealp{irions2017survey}), we might anticipate that, if individuals on the ends to the latent scale strongly agree with the ``positive" version of the statement, they will strongly disagree with its ``negative'' version, and vice versa.

To be more specific, let $y_{i,j}$ and $y_{i,j}'$ be the responses of individual $i$ to item $j$ under the `positive'' and ``negative'' versions, and $\theta_{i,j}(k)$ and $\theta_{i,j}'(k)$ the corresponding probabilities.  Consistency in the response patterns for item $j$ corresponds to requiring $\theta_{i,j}(k) = \theta_{i,j}'(K_j - k)$ for every $i$ and $k$.  In terms of the response functions, this implies that
\begin{equation}\label{eq:scalereversal}
\begin{aligned}
\Pr(y_{i,j} \leq k \mid \beta) &= \sum_{\ell=0}^{k} \theta_{i,j}(\ell) = \sum_{\ell=0}^{k} \theta_{i,j}'(K_j-\ell) = \sum_{\ell'=K_j-k}^{K_j} \theta_{i,j}'(\ell') \\
& = 1-\Pr(y_{i,j}' \leq K_j-k-1 \mid \beta)
\end{aligned}
\end{equation}
To illustrate the relationship, we present in Figure \ref{fig:flipped_scale} the implied response functions for the negative version of the item in Figure \ref{fig:bggum_cdf}.  As these figures demonstrate, one consequence of \eqref{eq:scalereversal} is that, if $\lim_{\beta\to-\infty} \Pr(y_{i,j} \leq k \mid \beta) = 1$, then (again, under the consistency assumption) $\lim_{\beta\to-\infty} \Pr(y'_{i,j} \leq k \mid \beta) = 0$.  Similarly if $\lim_{\beta\to\infty} \Pr(y_{i,j} \leq k \mid \beta) = 1$, then $\lim_{\beta\to\infty} \Pr(y'_{i,j} \leq k \mid \beta) = 0$.  Since GGUM can only generate response functions for which $\lim_{\beta\to-\infty} \Pr(y_{i,j} \leq k \mid \beta) = 1$ and $\lim_{\beta\to\infty} \Pr(y_{i,j} \leq k \mid \beta) = 1$, they can only properly capture the response function of ``positively'' worded items (e.g., those in the left panel of Figure \ref{fig:bggum_cdf}), and the model will necessarily lead to different estimates of the $\beta_i$s depending on how the item has been worded.  When there is enough substantive knowledge about the item to anticipate that individuals with extreme responses will disagree with it, this issue can be addressed by reversing the coding of the responses \textit{before} fitting the model. However, GGUMS have no way to automatically accommodate items with reversed scales without this prior knowledge and pre-processing step.}  

Another feature of the response functions implied by a GGUM that is worthwhile discussing is their symmetry  around $\mu_j$, i.e., the fact that 
$$
\Pr\left(y_{i,j} \le k \mid \beta_i = \mu_j + \Delta \right) = \Pr\left(y_{i,j} \le k \mid \beta_i = \mu_j - \Delta \right)
$$
for every $k \in \{ 0, 1, \ldots, K_j \}$ and $\Delta > 0$ (please see our supplementary materials).  Again, we argue that this is an undesirable feature of the model, as such strong constraint is unlikely to hold in practical applications (examples of this are presented Section \ref{sec:results}).  

To conclude, we note that, as far as we are aware, no representation of the GGUM in terms of (one or more) continuous latent variables is available.  This fact makes computation for GGUMs much more challenging than for GRMs. Furthermore, while $\beta_i$ and $\mu_j$ can still be interpreted as the latent trait and as a kind of difficulty parameter, the interpretation of $\alpha_j$ or the $\tau_{j,0}, \ldots, \tau_{j,K_j+1}$ is not straightforward (see \citealp{roberts2000general} for a detailed discussion).  

\section{An ordinal probit unfolding model based on random utilities}\label{sec:model}

We consider now an alternative unfolding model whose formulation relies on the random utility framework of \citet{mcfadden1973conditional}. We start by motivating the construction of the model and then move on to address the issues of the modeling of scale reversal in negatively worded items, as well as prior elicitation in the context of Bayesian inference.

To \editMark{motivate} our model, assume that subject $i$ chooses among the $K_j + 1$ levels in question $j$ according to a series of quadratic utilities 
\begin{align*}
    U(\beta_i, \psi_{j, k}) &= - \left| \beta_i - \psi_{j, k} \right|^2 + \epsilon_{i, j, k} , &  k=-K_j, -K_j+1, \ldots, 0 , \ldots K_j - 1, K_j ,
\end{align*}
where $\beta_i \in \real$ is the latent trait for individual $i$, the $\epsilon_{i,j,k}$'s are independent and identically distributed random shocks, and $\psi_{j,-K_j}, \psi_{j,-K_j +1}, \ldots, \psi_{j,K_j-1}, \psi_{j,K_j}$ are $2K_j + 1$ attraction points such that
\begin{align}\label{eq:ordconstraintpsi}
\psi_{j,-K_j} < \psi_{j,-K_j +1} < \cdots < \psi_{j,0} < \cdots \psi_{j,K_j-1} < \psi_{j,K_j}.    
\end{align}
\editMark{The use of quadratic utility functions in the context of random utility models has a long history, particularly in the context of spatial and item response theory model (e.g., see \citealp{JackmanMultidimensionalAnalysisRoll2001} and \citealp{fox2010bayesian}).  Conceptually, a key reason is that quadratic utility functions can be seen as second order (Taylor) approximation to other differentiable utility functions.  From a practical point of view, a major reason is that the simplifications that arise when computing differences between alternatives' utilities facilitate algorithm design. 
} As for the random shocks, we assume that the $\epsilon_{i,j,k}$s are independent and identically distributed from a standard Gaussian distribution.  Again, this choice leads to substantial computational advantages, but generalizations to other types of link functions are possible (e.g., see \citealp{lei2025logit}).

Let $\psi_{j,0}$ be the attraction point associated with $y_{i,j} = 0$ which, as in the case of GGUM's, corresponds to the strongest level of disagreement with the item.  Furthermore, let each pair $(\psi_{j,-k}, \psi_{j,k})$ be associated with the outcome $y_{i,j}=k$.  Then, we define
\begin{align}\label{eq:thetaprob}
    \theta_{i,j}(k) = \Pr\left( y_{i,j}=k \mid \beta_i \right) = 
    \begin{cases}
        \Pr\left(  {\displaystyle \argmax_{  m\in \{ -K_j, \ldots, K_j \}}}  U\left( \beta_i, \psi_{j,m} \right) = 0  \right)  & k = 0, \\
        \Pr\left(  {\displaystyle \argmax_{  m\in \{ -K_j, \ldots, K_j \}}}  U\left( \beta_i, \psi_{j,m} \right) \in \{ -k,k \}  \right) & k \ge 1 .
    \end{cases}
\end{align}
\editMark{The probability $\theta_{i,j}(k)$ can be expressed in terms of the cumulative distribution of a multivariate Gaussian distribution (please see the see the supplementary materials).}

\begin{figure}
    \centering
    \centering
    \begin{subfigure}[t]{0.45\textwidth}
        \includegraphics[width = \textwidth]{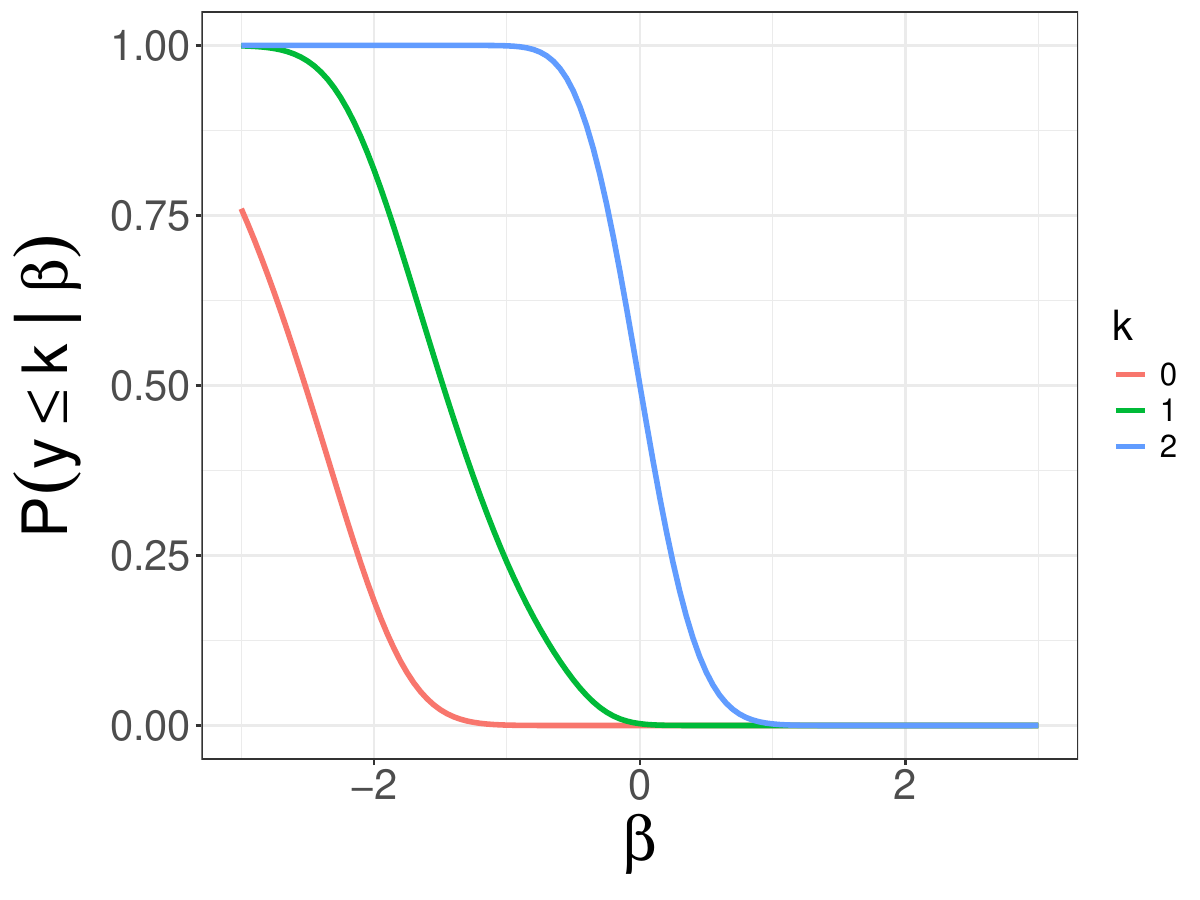}
        \centering
        \caption{
        }
        \label{fig:exampleresponses_mono}
    \end{subfigure}
    \begin{subfigure}[t]{0.45\textwidth}
        \includegraphics[width = \textwidth]{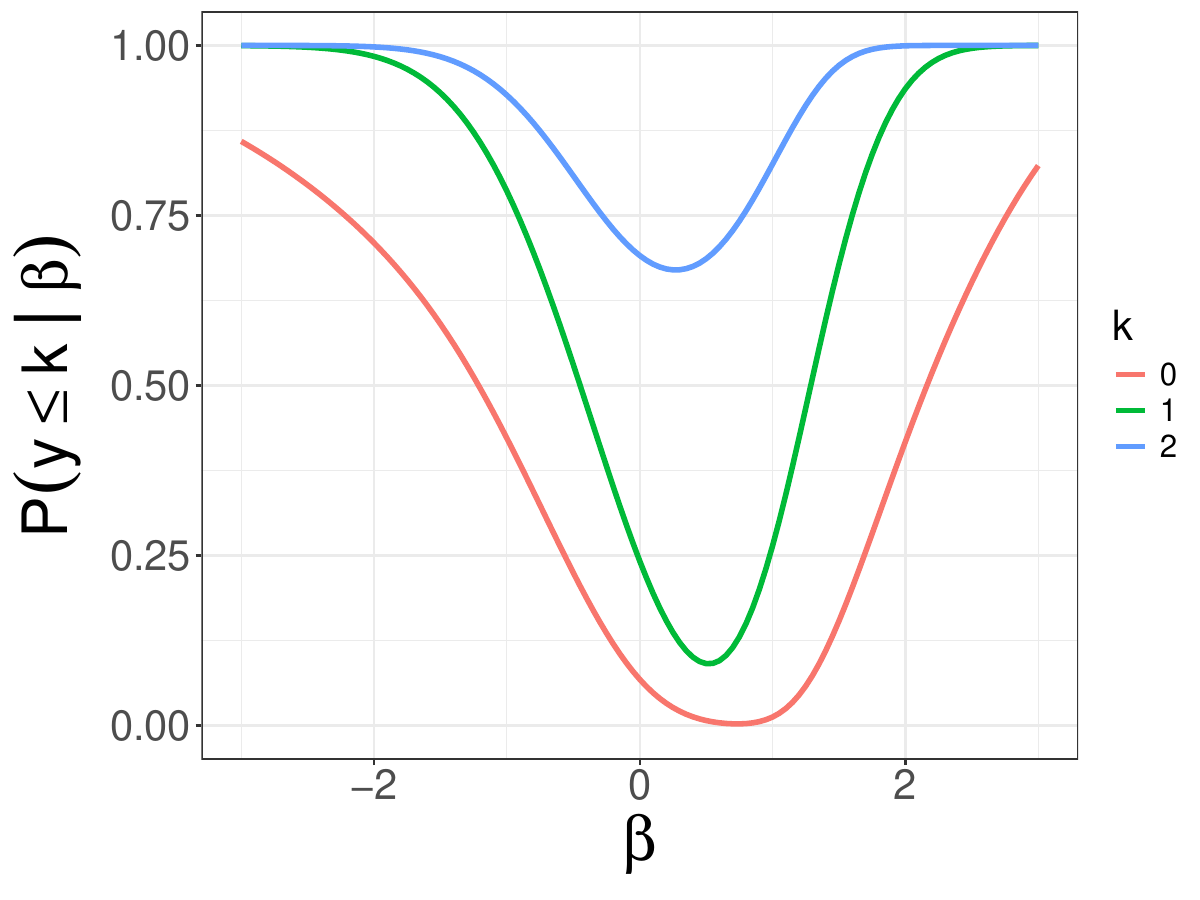}
        \caption{
        }
        \label{fig:exampleresponses_nonmono}
    \end{subfigure}
    \caption{Examples of response functions  for OPUM where $K_j=3$.  Panel (a) shows an example where the response functions are monotonic over the range of interest, which corresponds to $\v{\psi}_j = (-3.0, -2.0, -1.0, 1.0, 6.0, 7.0, 8.0)$ and (b) shows an example of non-monotonic response functions, which corresponds to $\v{\psi}_j = (-1,-0.65,-0.1, 0.1, 0.75, 1.75, 2.5)$.}
    \label{fig:exampleresponses}
\end{figure}

\editMark{This type of formulation, in which the probability of the response increases as the distance between the latent trait/ideal point and the attraction point decreases for a particular outcome, is the same basis on which spatial models are built \citep{EnelowHinichSpatialTheoryVoting1984}. Our formulation builds on traditional versions of this framework by introducing additional attraction points so that the level of agreement with a particular item can potentially increase both when the latent trait increases or decreases, combining the intuition behind the construction of GGUMs with that of ideal point and spatial models. In this way, the resulting model can capture non-monotonic forms for $\Pr(y_{i,j} \le k \mid \beta)$. Furthermore, as was the case for GGUMs, monotonic response functions over a range of interest can also be recovered through a careful selection of $\psi_{j,-k}$ and $\psi_{j,k}$s.  In fact, in the dichotomous case ($K_j=1$), letting either $\mu_{j,-1} \to -\infty$ or $\mu_{j,1} \to \infty$ leads back to the traditional probit IRT model \citep{fox2010bayesian}, which also corresponds to the ideal point model of \citet{JackmanMultidimensionalAnalysisRoll2001}. Figure \ref{fig:exampleresponses} shows examples of  parameter settings that lead to curves with either set of properties. These graphs also demonstrate that OPUM's response functions are not necessarily symmetric, which suggests that the model is more flexible than BGGUM.}

\editMark{An alternative construction of this spatial model} involves a latent variable representation in terms of a collection of $2K_j+1$ dimensional auxiliary vectors $\bfz_{i,j} \sim \textrm{N} \left( \bfeta_{i,j}, I_{2K_j+1} \right)$ where
\begin{align*}
    \bfeta_{i,j} &= (\eta_{i,j,-K_j}, \eta_{i,j,-K_j+1}, \ldots \eta_{i,j,-1}, 0, \eta_{i,j,1}, \ldots \eta_{i,j,K_j-1}, \eta_{i,j,K_j})' ,
\end{align*}
with $\eta_{i,j,k} = \alpha_{j,k} \left( \beta_i - \mu_{j,k} \right)$, $\alpha_{j,k} = 2\left( \psi_{j,k} - \psi_{j,0} \right)$, $\mu_{j,k} = \left( \psi_{j,k} + \psi_{j,0} \right)/2$, and $I_d$ denoting the $d \times d$ identity matrix.  Then, conditional on these auxiliary variables, $y_{i,j}=0$ if and only if $\argmax_{m} z_{i,j,m} = 0$, and $y_{i,j}=k$ if and only if $\argmax_{m} z_{i,j,m} \in \{-k,k\}$ for $k=1,\ldots, K_j$. 
\editMark{To see this, define 
\begin{align*}
U^{*}(\beta_i, \psi_{j,k}, \psi_{j,0}) &= U(\beta_i, \psi_{j,k}) - -\left| \beta_i - \psi_{j,0} \right|^2 \\
&= 2 \left( \psi_{j,k} - \psi_{j,0} \right) \left( \beta_i - \frac{\psi_{j,k} + \psi_{j,0}}{2} \right) + \epsilon_{i,j,k} \\
& = \alpha_{j,k} \left( \beta_i - \mu_{j,k} \right) + \epsilon_{i,j,k}  
\end{align*}
%
and note that
\begin{align*}
    \Pr\left(  { \displaystyle \argmax_{  m\in \{ -K_j, \ldots, K_j \} }}  U\left( \beta_i, \psi_{j,m} \right) = k  \right)
    &= \Pr\left(  { \displaystyle \argmax_{  m \in \{ -K_j, \ldots, K_j \} }}  U^{*}\left( \beta_i, \psi_{j,m} , \psi_{j,0} \right) = k  \right) \\
    &=\Pr\left(  { \displaystyle \argmax_{  m \in \{ -K_j, \ldots, K_j \} }}  z_{i,j,k} = k  \right)
\end{align*}

Our Ordinal Probit Unfolding Model (OPUM) corresponds to the probability model on the response matrix $\bfY$ under the $(\bfalpha, \bfmu)$ reparameterization of the model.  Working in this reparameterization allows for greater flexibility in terms of the shape of the response curves that can be generate by the model and facilitates computation.  Consistency with the order constrained in Equation \ref{eq:ordconstraintpsi} is achieved through careful prior elicitation (please see Section \ref{sec:prior}).
}

\subsection{\editMark{Scale reversals and negatively worded items}}\label{sec:sym}

The formulation we just described shares one unappealing feature with GGUMs:  the assumption that $y_{i,j}=0$ corresponds to the lowest level of agreement with the item statement, which  implies that the shape of the response functions is constrained so that $\lim_{\beta_i \to \infty} \Pr(y_{i,j}=0 \mid \beta_i) = \lim_{\beta_i \to -\infty} \Pr(y_{i,j}=0 \mid \beta_i) = 1$ (recall Figure \ref{fig:flipped_scale} and the associated discussion). 

We propose to address this shortcoming of the model by modifying it so that the likelihood contribution from each \textit{item} is defined as a mixture between those corresponding to the two possible ordering of the Likert scale, i.e., by setting 
\editMark{\begin{multline}\label{eq:symmetric}
    f\left(\bfY \mid \{\beta_i\}, \{ \mu_{j,k} \}, \{ \alpha_{j,k} \}\right) = \\
    \prod_{j=1}^{J}  \left\{ \lambda_j \prod_{i=1}^{I} \prod_{k=0}^{K_j}\left\{ \theta_{i,j}(k) \right\}^{\indFct{y_{i,j}=k}} + 
    (1 - \lambda_j) \prod_{i=1}^{I} \prod_{k=0}^{K_j} \left\{\theta_{i,j}(K_j-k) \right\}^{\indFct{y_{i,j}=k}}
    \right\} 
\end{multline}}
\editMark{for some $\lambda_j \in [0, 1]$.  Intuitively, the first component of the mixture assumes that response functions are v-shaped (similar to those shown in Figure \ref{fig:bggum_cdf}), while the second component assumes that they are n-shaped (similar to those in Figure \ref{fig:flipped_scale}).} 

A useful way to interpret this mixture model is by introducing an additional set of auxiliary variables $\zeta_1, \ldots, \zeta_J \in \{0,1\}$ such that \editMark{$\Pr(\zeta_j = 1) = \lambda_j$, $\Pr(\zeta_j = 0) = 1 - \lambda_j$,} and
\begin{multline}\label{eq:mixture_zeta}
    f\left(\bfY \mid \{\beta_i\}, \{ \mu_{j,k} \}, \{ \alpha_j \}, \{ \zeta_j \} \right) = \\
    \prod_{j=1}^{J} \prod_{k=0}^{K_j}  \prod_{i=1}^{I} \left\{ \theta_{i,j}(k) \right\}^{\indFct{y_{i,j}=k} \indFct{\zeta_{j}=0}}  \left\{\theta_{i,j}(K_j-k) \right\}^{\indFct{y_{i,j}=k} \indFct{\zeta_{j}=1}}
     .
\end{multline}
Integrating over the $\zeta_j$'s yields \eqref{eq:symmetric}. In agreement with the previous discussion, we can interpret $\zeta_j = 0$ as indicating that $y_{i,j} = 0$ corresponds to the highest level of disagreement with the concept embodied in the item (and therefore $y_{i,j} = K_j$ corresponds to the highest level of agreement), and $\zeta_{j} = 1$ as indicating that the wording of the question implies that $y_{i,j} = 0$ corresponds to the highest level of agreement (or the highest level of disagreement with the negated version of the item).  \editMark{Hence, which component of the mixture is appropriate for a given question depends on whether (in the language of Section \ref{sec:GRMGGUMreview}) the item has been worded ``positively'' or ``negatively''.
If the value of the $\zeta_j$'s can be safely determined from substantive knowledge about the items, the formulation can be simplified. In fact, note that the vanilla OPUM model corresponds to assuming $\zeta_j = 0$ for all $j=1,\ldots,J$, i.e., that all items have been worded positively.  On the other hand, when no substantive knowledge is available, treating the $\zeta_j$s as unknown parameters and inferring them from the data allows the model to learn the correct order of the Likert scale for each item.}

To conclude this section, it is worthwhile noting that, conceptually, a similar approach could be used in the context of GGUMs.  However, from a practical point of view, implementing this solution is much harder in that setting..

\subsection{Prior distributions}\label{sec:prior}
\label{ssec:prior}
Taking a Bayesian approach to inference for this model requires that we assign prior distributions to the parameters $\beta_1, \ldots, \beta_I$, $\bfalpha_1, \ldots \bfalpha_J$, $\bfmu_1, \ldots, \bfmu_J$ \editMark{ and, if unknown, $\zeta_1, \zeta_2, \ldots, \zeta_J$. For $j = 1, 2, \ldots, J$, we set the prior probability so that $\lambda_j = \textrm{Pr}(\zeta_j = 1) = 1/2$. While more general specifications (such as a Beta prior on $\lambda_j$) are possible and computationally straightforward, our numerical experiments suggest that the results are robust.} Furthermore, we adhere to common practice in the literature and assume that the latent traits $\beta_1, \ldots, \beta_I$ are independent and identically distributed according to a standard normal distribution.  The use of a normal prior helps address identifiability issues common in latent trait modeling (see Section \ref{sec:ident}) and simplifies computation (see Section \ref{sec:sampling}). 

As is also customary in the literature, we assume that the pairs of item-specific parameters $(\bfalpha_1, \bfmu_1), \ldots, (\bfalpha_J, \bfmu_J)$ are independent and identically distributed (the so-called local independence assumption).  Beyond that, choosing the specific form of the joint prior for each pair requires some care.  In particular, note that such prior should retain some consistency with the order constraint in \eqref{eq:ordconstraintpsi}, which in turn implies order constraints of the $\alpha_{j,k}$s and on the $\mu_{j,k}$s.  To account for this we work with a truncated Gaussian prior
\begin{align*}
    \left(\bfalpha_j, \bfmu_j\right) \sim \textrm{N}\left( \bfvarphi_j , \bfSigma_j \right) \indFct{\bfalpha_j \in \Omega_j}
\end{align*}
where
\begin{align*}
    \Omega_j = \left\{ \bfalpha_j : \alpha_{j,-K_j} < \alpha_{j,-K_j+1} < \cdots < \alpha_{-1} < 0 < \alpha_1 < \cdots < \alpha_{j,K_j-1} < \alpha_{j,K_j}
    \right\} ,
\end{align*}
$\bfvarphi_j' = ( \mathbf{0}_{2K_j}', \bfupsilon_j')$, $\mathbf{0}_{d}$ is a vector of zeros of dimension $d$ as before, $\bfupsilon_j$ is a vector of length $2K_j$ such that $\upsilon_{j,k} = - \upsilon_{j,2K_j - k + 1}$, $\bfSigma_j = \Bdiag \left\{ \kappa_j^2 I_{2K_j} , \omega_j^2 I_{2K_j} \right\}$, and $I_d$ again stands for the $d \times d$ identity matrix.  \editMark{Note that truncating the Gaussian distribution over the set $\Omega_j$ and setting $\upsilon_{j,k} = - \upsilon_{j,2K_j - k + 1}$ are both consistent with \eqref{eq:ordconstraintpsi}.}

Specific values for $\bfupsilon_j$, $\kappa_j^2$ and $\omega_j^2$ can be elicited by examining the \textit{a priori} shape of the implied response functions, which are identifiable and interpretable.  In the absence of strong prior information, we recommend aiming for response functions that are, \textit{a priori}, relatively constant and that assign all categories similar probabilities.  A concrete example of hyperparameter choices is presented in Section \ref{sec:results}.

\editMark{\subsection{Parameter identifiability}\label{sec:ident}

Fixing the mean and variance of the prior on the latent traits $\beta_1, \ldots, \beta_I$ allows us to address identifiability issues related to the invariance of the likelihood to translations and rescalings of the underlying latent scale (recall Section \ref{ssec:prior}). On the other hand, we address invariance to reflections/rotations by post-processing the samples of the posterior distribution to fix the sign of the ideal point of one pre-selected $\beta_i$ (e.g., see \cite{bafumi2005practical}).

Another form of (potential) lack of identifiability in our model relates to the mixture model formulation in Section \ref{sec:sym} and the possibility of label switching.  This not an issue in our case because the two components corresponds to distinct probabilistic models with unique properties (e.g., different limiting behavior as $\beta \to \pm \infty$).}

\section{Computation}\label{sec:sampling}

We propose a Markov chain Monte Carlo (MCMC) algorithm to generate approximate samples from the posterior distribution of our model.  As is customary, empirical summaries of these samples can then be used to approximate point and interval estimators for various parameters of interest.

The algorithm we proposes relies mostly on Gibbs-type steps (where the full conditional posterior distributions belong to well-known families and can therefore be sampled directly), supplemented with carefully chosen Metropolis-Hastings steps to improve mixing. To accomplish this, we augment the model with two different sets of auxiliary variables.  

The first set of auxiliary variables are a set of $I \times J$ vectors $\bfz_{i,j}$ of dimension $2K_j+1$, as defined in Section \ref{sec:model}.  Conditional on these random vectors, the full conditional distribution for $(\bfalpha_j,\bfdelta_j)$ corresponds to a (truncated) multivariate Gaussian distributions.  Conversely, conditional on the remaining parameters, each $z_{i,j,k}$ also follows an appropriately truncated Gaussian distribution.  This portion of the algorithm is reminiscent of the approach introduced in  \citet{albert1993bayesian} for sampling from the posterior distribution of probit regression models.

The second set of auxiliary variables correspond to the indicators $\zeta_1, \ldots, \zeta_J$ introduced in Section \ref{sec:sym}.  We sample each $\zeta_j$ using a Metropolis-Hastings step in which the proposals are generated by first flipping the current value of $\zeta_j$ and then by generating new values of the item-specific parameters $(\bfalpha_j, \bfdelta_j)$ from their prior distribution.  The acceptance probability for such a proposal is given by $\min\left\{ 1, \Delta_j\right)$,
where  
$$
\Delta_j = \begin{cases}
   \prod_{k=0}^{K_j} \prod_{i=1}^{I} \left\{\frac{\theta_{i,j}^{(p)}(K_j-k)}{\theta_{i,j}^{(c)}(k)}\right\}^{\indFct{y_{i,j}=k}} & 
   \zeta_j^{(c)} = 0 , \\
   \prod_{k=0}^{K_j} \prod_{i=1}^{I} \left\{\frac{\theta_{i,j}^{(p)}(k)}{\theta_{i,j}^{(c)}(K_j-k)}\right\}^{\indFct{y_{i,j}=k}} & \zeta_j^{(c)} = 1 ,
\end{cases}
$$
$\theta_{i,j}(k)$ is given by equation \eqref{eq:thetaprob}, and the $(c)$ and $(p)$ superscripts indicate the current and proposed values of the parameters. Because the computing $\theta_{i,j}(k)$ is costly, we perform this MCMC step only every 50 iterations of the chain. \editMark{Note that using a lower (but constant over time) frequency for some steps does not invalidate the MCMC algorithm.}

A consequence of the lack of monotonicity of the response functions is that the posterior distribution of $\beta_i$ can be multimodal. To account for this fact, we incorporate in our MCMC algorithm an additional Metropolis-Hastings step in which a new value of the latent trait $\beta_i$ is proposed from a Gaussian distribution centered on the negative of its current value, i.e., $\beta_i^{(p)} \sim \textrm{N}\left( -\beta_i^{(c)}, \lambda^2 \right)$ while keeping all other parameters constant.  The associated acceptance probability becomes 
$$
\min \left\{ 1 ,  \prod_{j=1}^{J} \prod_{k=0}^{K_j} \left[ \left( \frac{\theta^{(p)}_{i,j}(k)}{\theta^{(c)}_{i,j}(k)} \right)^{\indFct{\zeta_j=0}} \left( \frac{\theta^{(p)}_{i,j}(K_j-k)}{\theta^{(c)}_{i,j}(K_j-k)} \right)^{\indFct{\zeta_j=1}} \right]^{\indFct{y_{i,j}=k}} \right\} .
$$
As before, because the computing $\theta_{i,j}(k)$ is costly, we perform this MCMC step only every 50 iterations of the chain.

An implementation of this algorithm is available at Githhub,
along with instructions for replicating the results discussed in the next seciton of the manuscript.

\section{Illustrations}\label{sec:results}

\subsection{A real data example: immigration policy preferences in the U.S.}\label{sec:realdata}

\citet{duck2022ends} consider data from a survey consisting of $J=10$ questions related to immigration policy in the U.S.\ Respondents were asked to give their preferences on an ordinal scale from 0 (strongly disagree) to 4 (strongly agree) for all questions. By construction, a small number of items were ``partisan'', i.e., expected to trigger strong disagreement from individuals on one side of the liberal-conservative ideological spectrum and strong agreement from individuals on the other side, while the remaining items were expected to lead to  disagreement from individuals with ideological preferences on either extreme of the ideological spectrum. Respondents who failed attention checks or answered only 1's or 5's on all questions were dropped from the survey, leading to a final sample that includes $I=2,621$ individuals.  The level of non-response is minimal:  only eight out of the $26,210$ answers are  missing.  Hence, for the purpose of our analysis, we treat missing values as missing completely at random. 
In addition to the 10 substantive questions, the survey asked respondents about their party and ideological leanings.  We do not use such information to estimate subjects' preferences, but we do use it as a \textit{post-hoc} check of the estimates generated by the various models.

\begin{figure}[!tb]
    \centering
    \includegraphics[width = \textwidth]{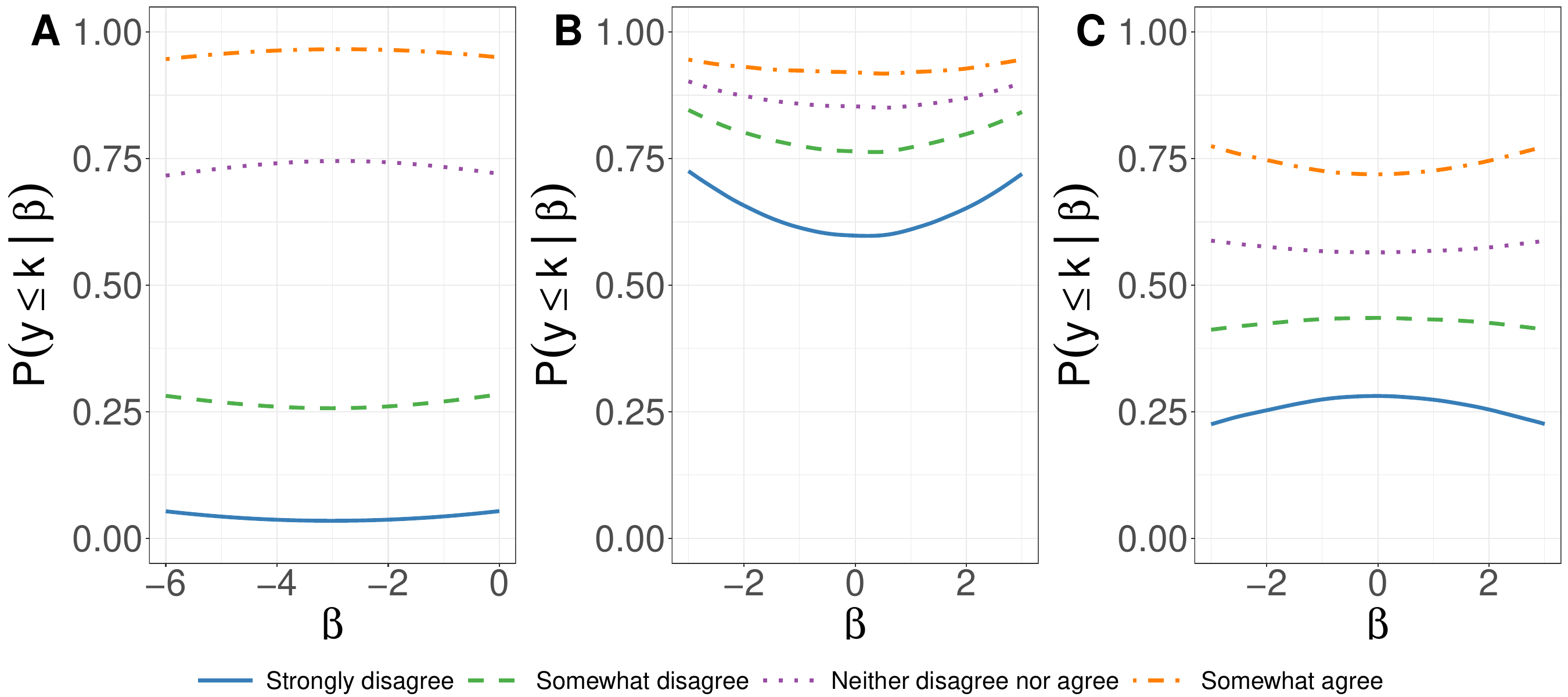}
    \caption{Prior mean response functions for $K_j=4$. Panel (a) corresponds to a BGRM with  $\alpha_j \sim N(0, 0.5)$ and a $\bftau_j \sim N((-4, -2, 2, 4)', 9 I_{4}) \indFct{\tau_{j, 0} < \tau_{j, 1} < \tau_{j, 2} < \tau_{j, 3}}$. Panel (b) corresponds to a BGGUM with $\alpha_j \sim \textrm{Beta}_{(0.25, 4)}(1.5, 1.5)$, $\mu_j \sim \textrm{Beta}_{(-5, 5)}(2, 2)$ and $ \tau_{j, l} \sim \textrm{Beta}_{(-6, 6)}(2, 2)$.  Panel (c) corresponds to an OPUM with $\kappa^2_j = 10$, $\omega^2_j = 25$, $\upsilon_{j, -k} = -4 - 1.5 k$ and $\upsilon_{j, k} = 4 + 1.5 k$.}
    \label{fig:bggum_prior_response_curve}
\end{figure}

We analyze these data using three different models: (a) a Bayesian version of the GRM model discussed in Section \ref{sec:GRMGGUMreview} (called BGRM in the sequel), (b) the Bayesian version of GGUM from Section \ref{sec:GRMGGUMreview} developed in \citet{duck2022ends} (referred to as BGGUM in the sequel), and (c) our OPUM model.  Samples from the posterior distribution under the BGRM were obtained using \texttt{RStan} \parencite{StanDevelopmentTeam.StanCoreLibrary2018}. The priors for the model parameters correspond to $\beta_i \sim \mathrm{N}(0,1)$, $\alpha_j \sim \mathrm{N}(0,3)$ and $\bftau_j \sim N((-4, -2, 2, 4)', 9 I_{4}) \indFct{\tau_{j, 0} < \tau_{j, 1} < \tau_{j, 2} < \tau_{j, 3}}$, independently across all $i$, $j$ and $k$.  Posterior summaries were generated using 10,000 samples obtained after a burn-in period of 1,000 iterations.  Results for the BGGUM were generated using the replication code in \citet{duck2022ends}, including their choice of priors.\footnote{We detected and corrected a mistake in the data pre-processing step of \citet{duck2022ends} replication code in which the order of two categories was transposed.  Because of this, the results under their model that we discuss here are somewhat different from those reported in their manuscript.}  Again, the prior on the latent traits corresponds to a standard Gaussian distribution, $\beta_i \sim \mathrm{N}(0,1)$.  On the other hand, the priors on $\alpha_j$, $\mu_j$ and $\tau_{j,k}$ correspond to shifted and scaled beta distributions, $\alpha_j \sim \textrm{Beta}_{(0.25, 4)}(1.5, 1.5)$, $\mu_j \sim \textrm{Beta}_{(-5, 5)}(2, 2)$ and $ \tau_{j, l} \sim \textrm{Beta}_{(-6, 6)}(2, 2)$.  See \citet{duck2022ends} for additional details.  In this case, inferences are based on 10,000 samples obtained after a burn-in period of 5,000 iterations.  
Finally, for OPUM, we employ the algorithm outlined in Section \ref{sec:sampling} with $\kappa^2_j = 100$, $\omega^2_j = 25$, $\upsilon_{j, k} = -4 + 1.5 k$ for $k < 0$ and $\upsilon_{j, k} = 4 + 1.5 k$ for $k > 0$.  A total of 10,000 samples obtained after a burn-in period of 200,000 samples and thinning every 10 samples were used for inference.  \editMark{Convergence of the three MCMC algorithms was monitored using the method of \citet{GelmanRubinInferenceIterativeSimulation1992} applied to four independent chains started at over-dispersed points. Convergence and mixing metrics, including $\hat{R}$ and equivalent sample sizes, are available in the supplementary materials.  As would be expected, execution times vary substantially, both because of the use of different computational platforms and because of sustantial differences in model complexity.}

\editMark{Figure \ref{fig:bggum_prior_response_curve} shows the implied prior mean response functions associated with the Bayesian BGRM, BGGUM and OPUM.  While not identical, the priors for BGRM and OPUM are similar in that the graphs show symmetry around the horizontal line at 1/2.  In contrast, the prior for BGGUM (which, as mentioned before, match those used used in \citealp{duck2022ends}) does not show such symmetry and implies an assumption that most individuals will disagree with the survey items. In fact, this is a structural shortcoming of BGGUMs, as it is impossible to get a symmetric graph no matter how the prior is selected.  This shortcoming is in part a byproduct of the fact that BGGUM cannot accommodate scale reversals due to negatively worded items (recall the discussions in Sections \ref{sec:GRMGGUMreview} and \ref{sec:sym}).  In any case, our sensitivity analysis indicates that results are robust to prior choices for all three models.}


\begin{figure}[!tb]
    \centering
    \includegraphics[width = \textwidth]{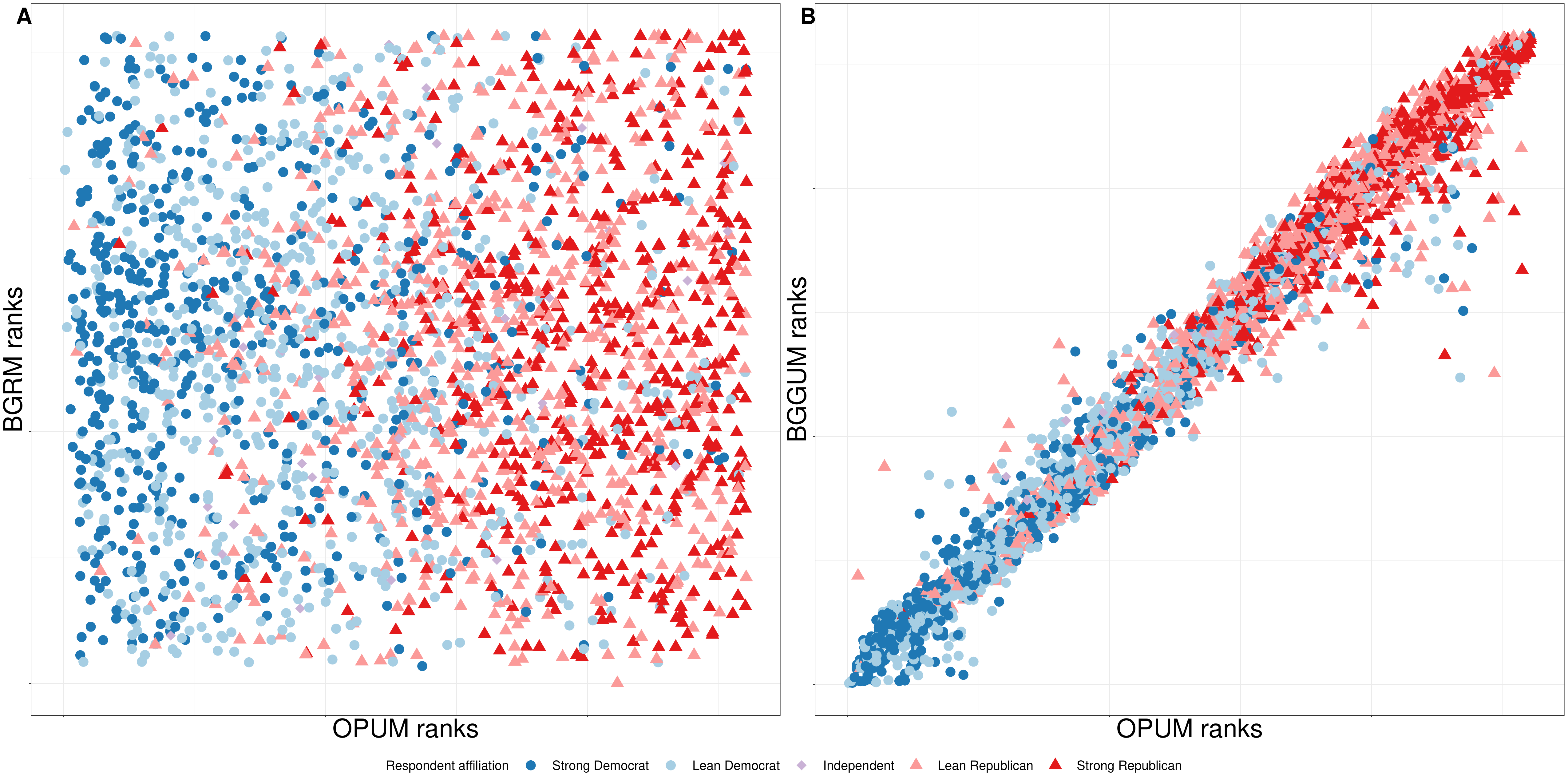}     
    \caption{Comparison of the posterior median ranks of individual preferences.  Panel (a) compares the ranks from BGRM from those from OPUM, while panel (b) compares the ranks from BGGUM to those of OPUM. Symbols are used to signal self-reported party affiliation (triangles for Republicans and circles for Democrats), while color saturation indicates the strength of the affiliation.}
    \label{fig:post_mean_ranks}
\end{figure}

We start our analysis by comparing the median ranks of the $\beta_i$'s for each individual across all three models (see Figure \ref{fig:post_mean_ranks}).  Panel (b) shows that the two models that allow for non-monotonic response functions (OPUM and BGGUM) are roughly in agreement, leading to estimates of preferences that tend to separate those that self-identify as Democrats (which appear concentrated on the bottom left quadrant of the plot) from those that self-identify as Republicans (which are concentrated on the top right quadrant). There are, nonetheless, some important differences across these two models when it comes to the estimates for specific individuals.  In particular, we can see a small group of subjects for whom OPUM's estimated preferences are more extreme than those estimated by BGGUM.  In contrast, panel (a) shows that BGRM leads to a very different ranking of individuals, one in which the rankings on the latent trait seem to be uncorrelated with self-reported party affiliation.

\editMark{As a first metric of model performance, }we use the Watanabe-Akaike Information Criteria (WAIC, \citealp{watanabe2010asymptotic, watanabe2013widely, gelman2014understanding}) to provide a more formal comparison of complexity-adjusted fit among the three models. WAIC is defined as following:
\begin{multline}\label{eq:lWAIC}
    \textrm{WAIC} = \editMark{-2}\log\left( \textrm{E}_{\textrm{post}} \left\{ \prod_{i=1}^I \prod_{j=1}^J \prod_{k=0}^{K_j}
    \left\{ \theta_{i,j}(k) \right\}^{\indFct{y_{i,j} = k}} \right\} \right)
    \\
    \editMark{+ 2} \sum_{i=1}^{I} \textrm{var}_{\textrm{post}}\left\{ 
     \left[\sum_{i = 1}^{I} \sum_{j = 1}^{J} \sum_{k=0}^{K_j} \indFct{(y_{i,j} = k} \log \theta_{i,j}(k) \right] \right\},
\end{multline}
where, as before, $\theta_{i,j}(k)$ represents the probability that individual $i$ chooses response $k$ on question $j$ under the given model. The first term in the WAIC formula measures goodness-of-fit to the data, while the second term acts as a complexity penalty.  Hence, \editMark{lower} values of WAIC indicate that the model provides a better complexity-adjusted fit to the data.  For the dataset we discuss in this section, 
we have $\textrm{WAIC}_{BGRM} = 77,064$, $\textrm{WAIC}_{BGGUM} = 69,430.76$, and $\textrm{WAIC}_{OPUM} = 68,590.54$, 
indicating that OPUM outperforms both alternatives under consideration, and confirming that BGRM provides a very poor fit to the data.

\editMark{In addition to comparing WAIC across the models, we also carry out posterior predictive checks for all 3 models.  These checks compare a sample of the posterior predictive distribution $p(\bfY^{*} \mid \bfY)$ from each of the models against the observed sample $\bfY$.  The first such comparison is based on the Wasserstein distance between the marginal predictive distribution of the responses for each item under the models and the observed distribution of responses under the data\footnote{We use the \texttt{R} package \texttt{transport} to compute the Wasserstein distances}.  Note that, because $K_j=4$ for all $j$, the maximum Wasserstein distance is 4 for all items.
%
%
Figure \ref{fig:post_pred_wasserstein_comps} shows the posterior distributions of the distances are, in all cases, quite low, suggesting that all models provide reasonable reconstructions. Nonetheless it is clear that BGRM leads to slightly larger distances than BGGUM and OPUM, indicating a slightly poorer fit. This is consistent with the WAIC results. The two unfolding models perform very similarly under this metric.


\begin{figure}
    \centering
    \begin{subfigure}[t]{0.45\textwidth}
        \includegraphics[width = 0.9\textwidth]{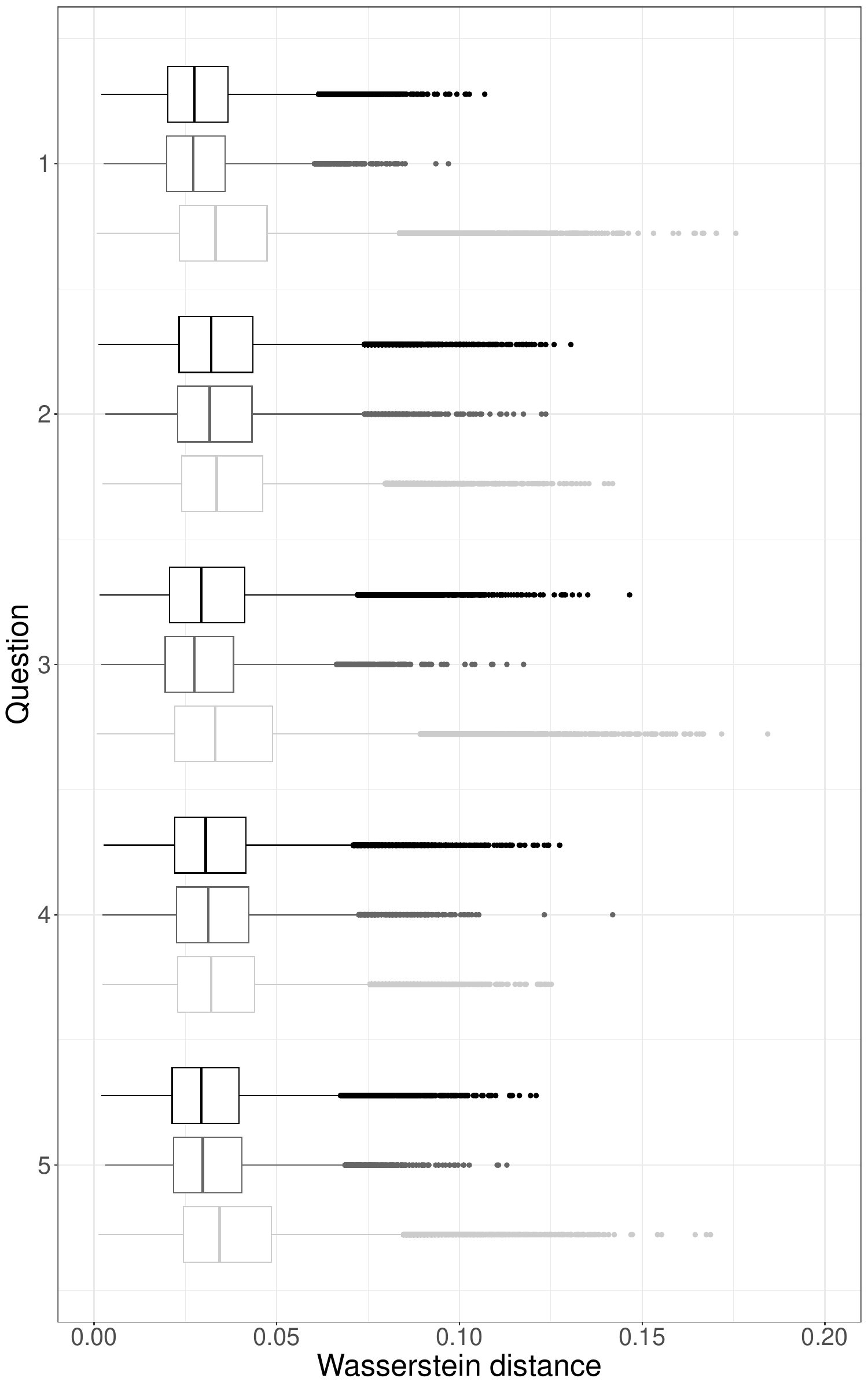}
        \centering
        %
    \end{subfigure}
    \begin{subfigure}[t]{0.45\textwidth}
        \includegraphics[width = 0.9\textwidth]{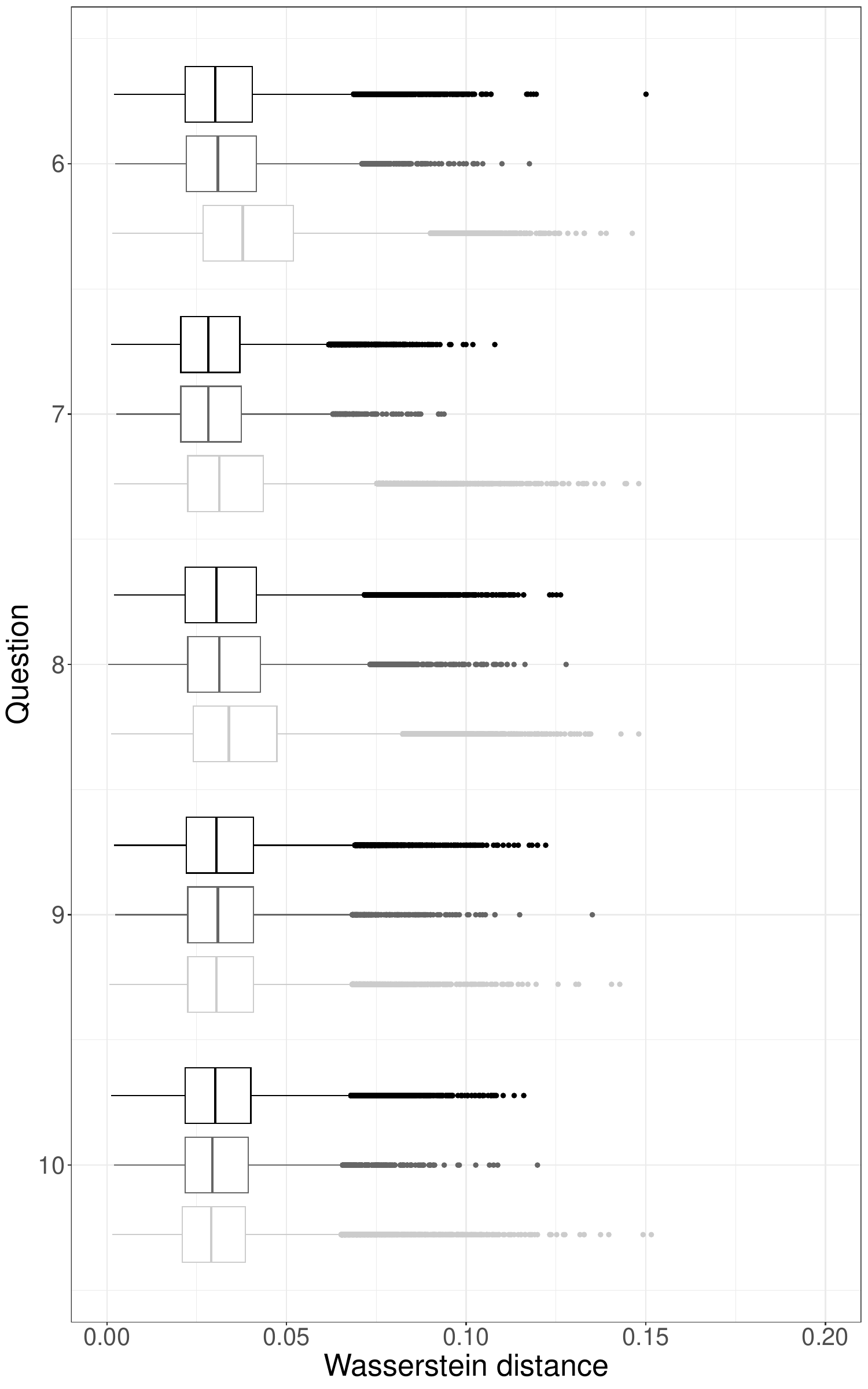}
        %
    \end{subfigure}
    \caption{\editMark{Wasserstein distance between the observed distribution of responses and samples of the marginal predictive distribution of the responses for each item under OPUM (black), BGGUM (dark grey) and BGRM (light grey). The left plots are for questions 1 through 5 and the rights plots are for questions 6 through 10.}}
    \label{fig:post_pred_wasserstein_comps}
\end{figure}


Our second predictive check uses ``second-order" metrics (the correlation among the responses from pairs of questions) to provide a more nuanced comparison between the predicted and actual response. More specifically, we compute the Spearman rank and Kendall-tau correlation for each posterior sample and pair of questions and check whether the observed correlation is within the 95\% credible intervals. For Spearman's rank correlation, the coverage of this intervals is 29 (OPUM), 24 (BGGUM) and 24 (BGRM) pairs of the total 45.  In the case of Kendall's tau, the coverage is 27 (OPUM), 22 (BGGUM) and 24 (BGRM).  Again, this suggests that OPUM outperforms BGGUM and BGRM.
}

\begin{figure}[!tb]
    \centering
    \includegraphics[width = \textwidth]{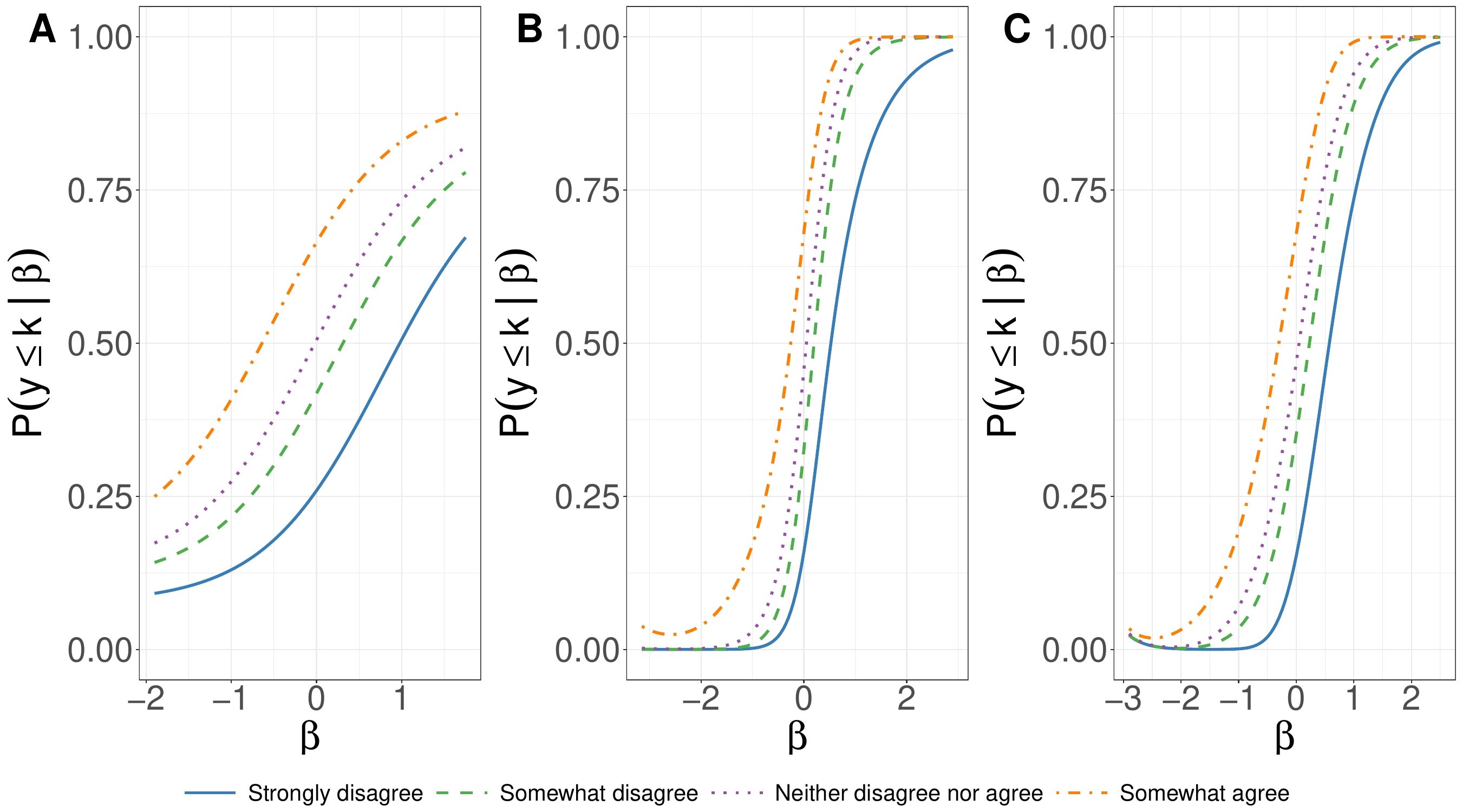}
    \caption{Posterior mean of the response functions for Question 3 estimated by each of BGRM (panel A), GGUM (panel B) and OPUM (panel C).  The horizontal axis of each panel captures the range of of values of the latent traits estimated by each model.}
    \label{fig:question_3_response_curve}
\end{figure}

A detailed analysis of the response curves associated with various items can provide useful insights into the differences in performance across models. We begin with Question 3, ``The U.S.\ does not need a wall along the entire U.S.-Mexican border.'' Given the salience of the issue and how clearly the parties were split on it at the time the survey was conducted, it would be expected that the models recover monotonically increasing functions (as individuals with Democrat leanings would be expected to largely agree with the statement, while individuals with Republican leanings would be expected to largely disagree).  As seen in Figure \ref{fig:question_3_response_curve}, this is indeed the case. The main difference across the estimates generated by the various models is the fact that the response functions for both GGUM and OPUM seem to be highly discriminant (i.e., have a steeper slope) but present minor inflection points at the lower end of the latent scale, while the response functions for BGRM seem to have less discrimination power.  These patterns seem to be driven by how each of the models accommodates a small number of respondents with extreme negative latent traits who report disagreeing with the statement.  In any case, the response functions generated by GGUM and OPUM would seem more in line with prior expectations about the shape of these particular response functions.

\begin{figure}[!tb]
    \centering
    \includegraphics[width = \textwidth]{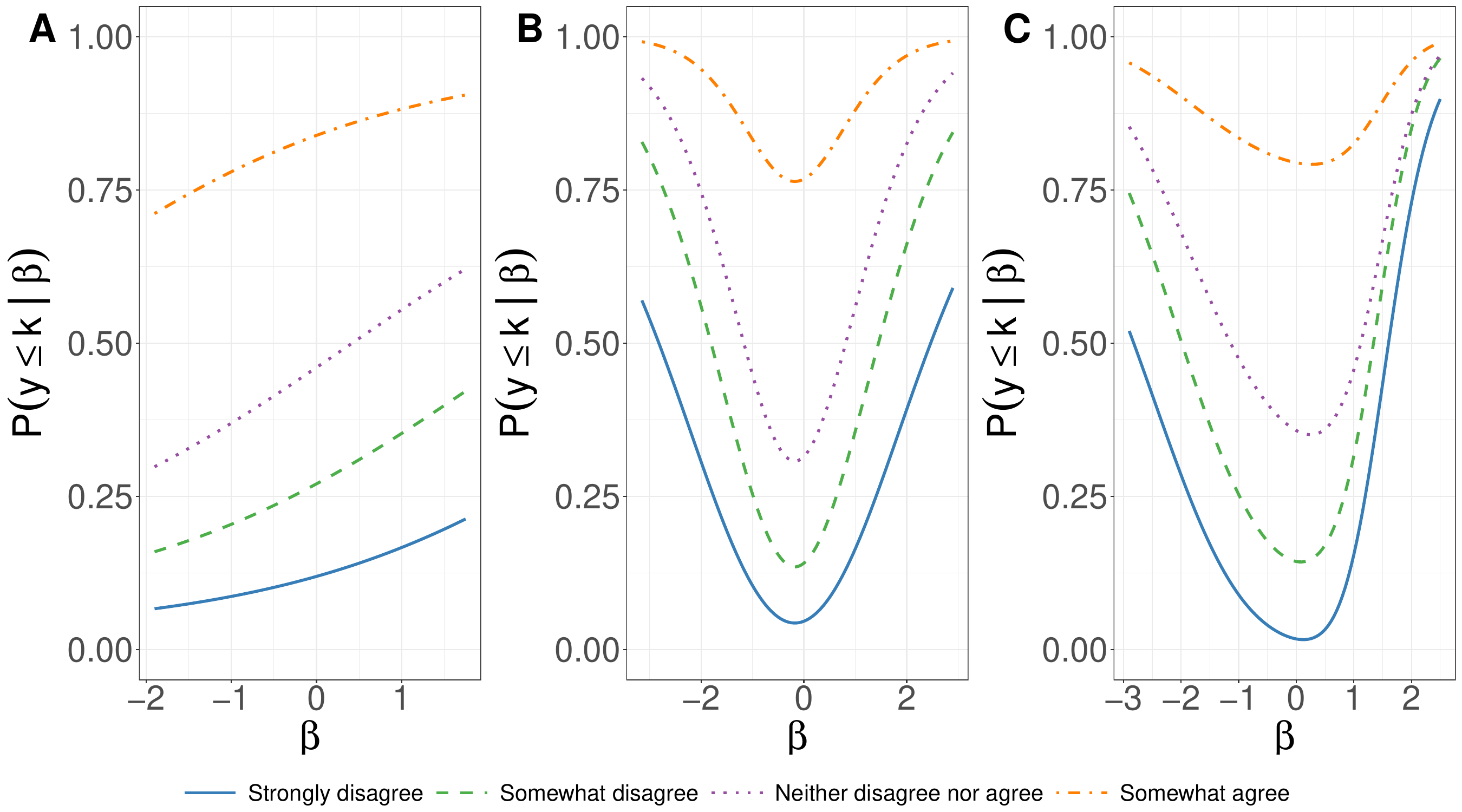}
    %
    %
    %
    %
    \caption{Posterior mean of the response functions for Question 2 estimated by each of BGRM (panel A), GGUM (panel B) and OPUM (panel C). The horizontal axis of each panel captures the range of of values of the latent traits estimated by each model.}
    \label{fig:question_2_response_curve}
\end{figure}

Next, we examine the response functions associated with Question 2, ``There should be a way for undocumented immigrants currently living in the U.S. to stay in the country legally, but only if certain requirements are met like learning English and paying a significant fine''.  The inclusion of the conditional statement in the wording of the item can be expected to lead to non-monotonic response functions, with Republicans potentially disagreeing with the existence of a path to legal status, and Democrats potentially agreeing to the path to legalization but objecting to some or all of the conditions required.  This expectation is well captured by the estimates generated by both GGUM and OPUM, which are non-monotonic (see Figure \ref{fig:question_2_response_curve}). Nonetheless, there are obvious and important differences between the two models.  GGUM estimates response function that are symmetric around zero, indicating that individuals at both ends of the ideological spectrum can be expected to disagree with the item in about equal measure.  In contrast, OPUM's estimates of the response functions indicate that the disagreement from extreme Democrats is much milder than the disagreement from extreme Republicans.  Again, given the nature of the discourse around immigration in the U.S.\ at the time of the survey, OPUM's response functions seem more appropriate. Finally, we note that BGRM (which, as discussed before, cannot accommodate non-monotonic responses) yields estimates of response functions that are increasing (which seems appropriate within the limitations of the model) but have limited discrimination power. 

\begin{figure}[!tb]
    \centering
    \includegraphics[width = \textwidth]{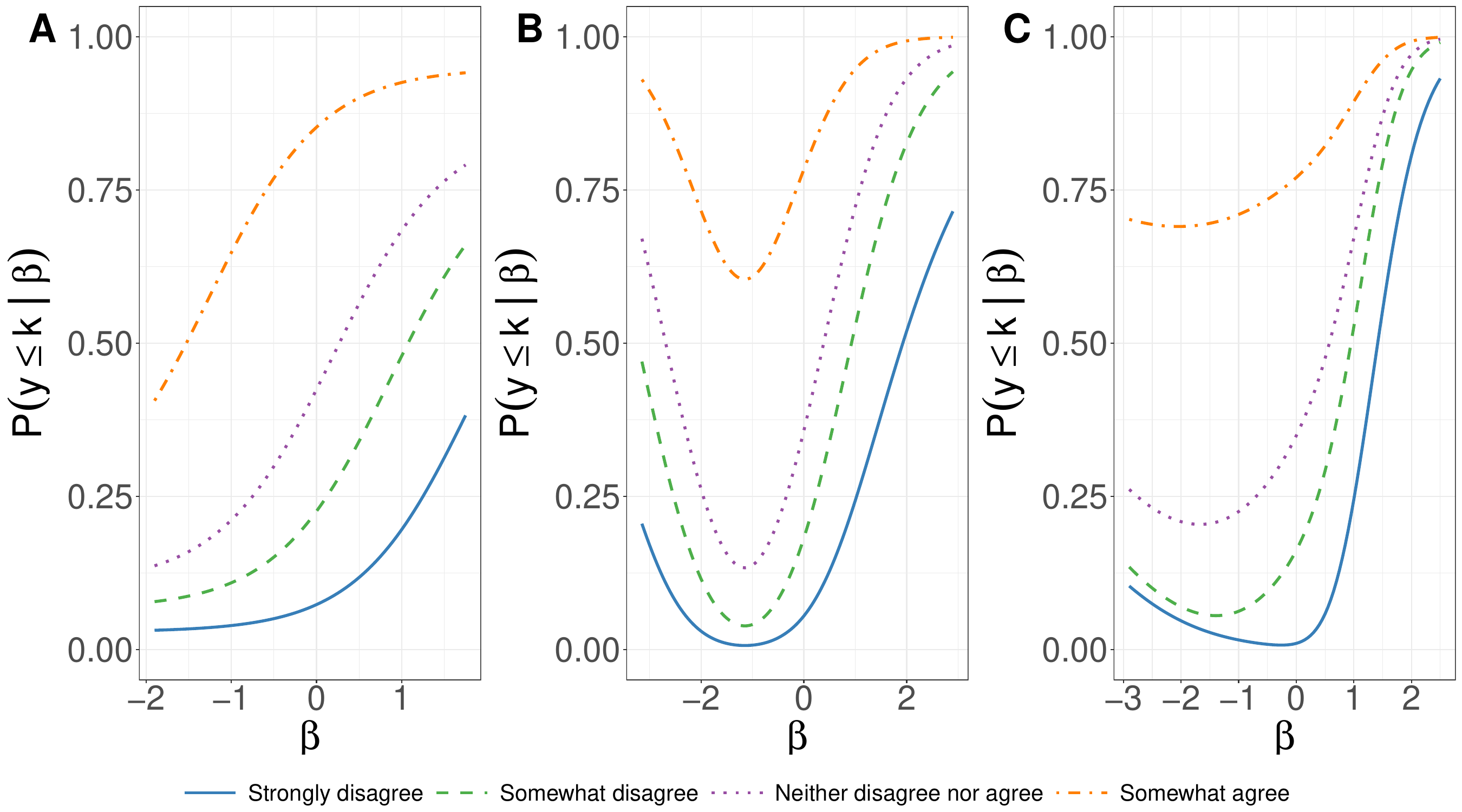}
    \caption{Posterior mean of the response functions for Question 6 estimated by each of BGRM (panel A), GGUM (panel B) and OPUM (panel C). The horizontal axis of each panel captures the range of of values of the latent traits estimated by each model.}
    \label{fig:question_6_response_curve}
\end{figure}

Similar patterns can be observed in the response functions associated with Question 6, ``The U.S. Congress should reach a compromise on immigration policy to allow in more immigrants but also improve enforcement.'' (see Figure \ref{fig:question_6_response_curve}). Again, both GGUM and OPUM yield estimates of the response functions that are non-monotonic.  However, the differences in this case are more striking, suggesting that the symmetry constraint associated with BGGUM might help explain the lower WAIC values we discussed before. 

\begin{figure}[!tb]
    \centering
    \includegraphics[width = \textwidth]{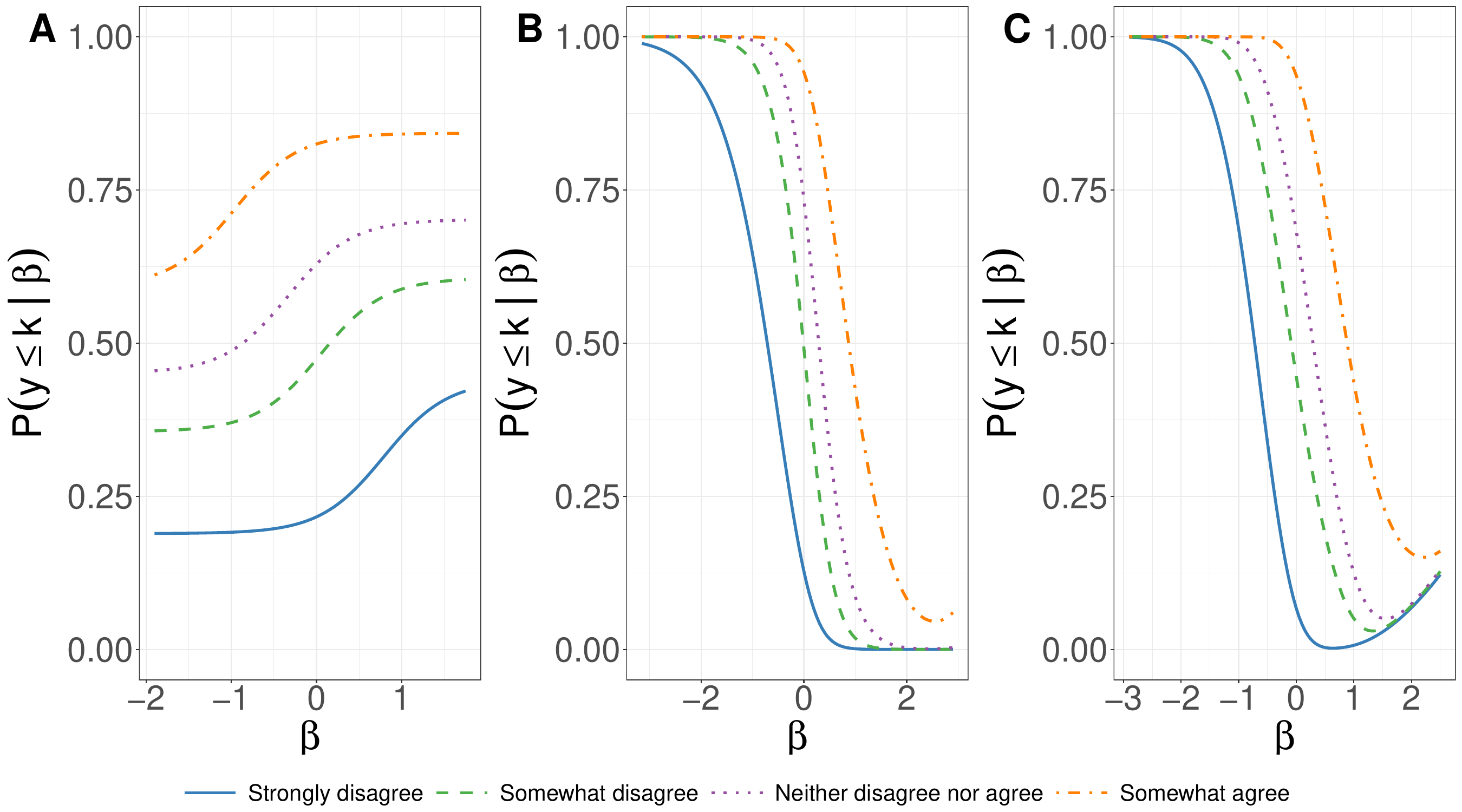}
    \caption{Posterior mean of the response functions for Question 1 estimated by each of BGRM (panel A), GGUM (panel B) and OPUM (panel C). The horizontal axis of each panel captures the range of of values of the latent traits estimated by each model.}
    \label{fig:question_1_response_curve}
\end{figure}

Finally, we examine the estimated response functions for Question 1 in the survey, ``All undocumented immigrants currently living in the U.S. should be required to return to their home country'' (see Figure \ref{fig:question_1_response_curve}).  This is, again, an item for which a monotonically decreasing response function could be expected, with Republicans strongly agreeing and Democrats strongly disagreeing. The estimates generated by both GGUM and OPUM are (for the most part) consistent with this expectation, although we must note that, somewhat surprisingly, OPUM identifies a small percentage (around 10\%) of extreme-right respondents who seem to disagree with the statement.  In contrast BGRM yields monotonically \textit{increasing} but only weakly discriminant response functions, a result that is perplexing and is probably driven by overall poor estimates of the latent traits (recall Figure \ref{fig:post_mean_ranks}).

\begin{figure}[!t]
    \centering
    \includegraphics[width=0.8\linewidth]{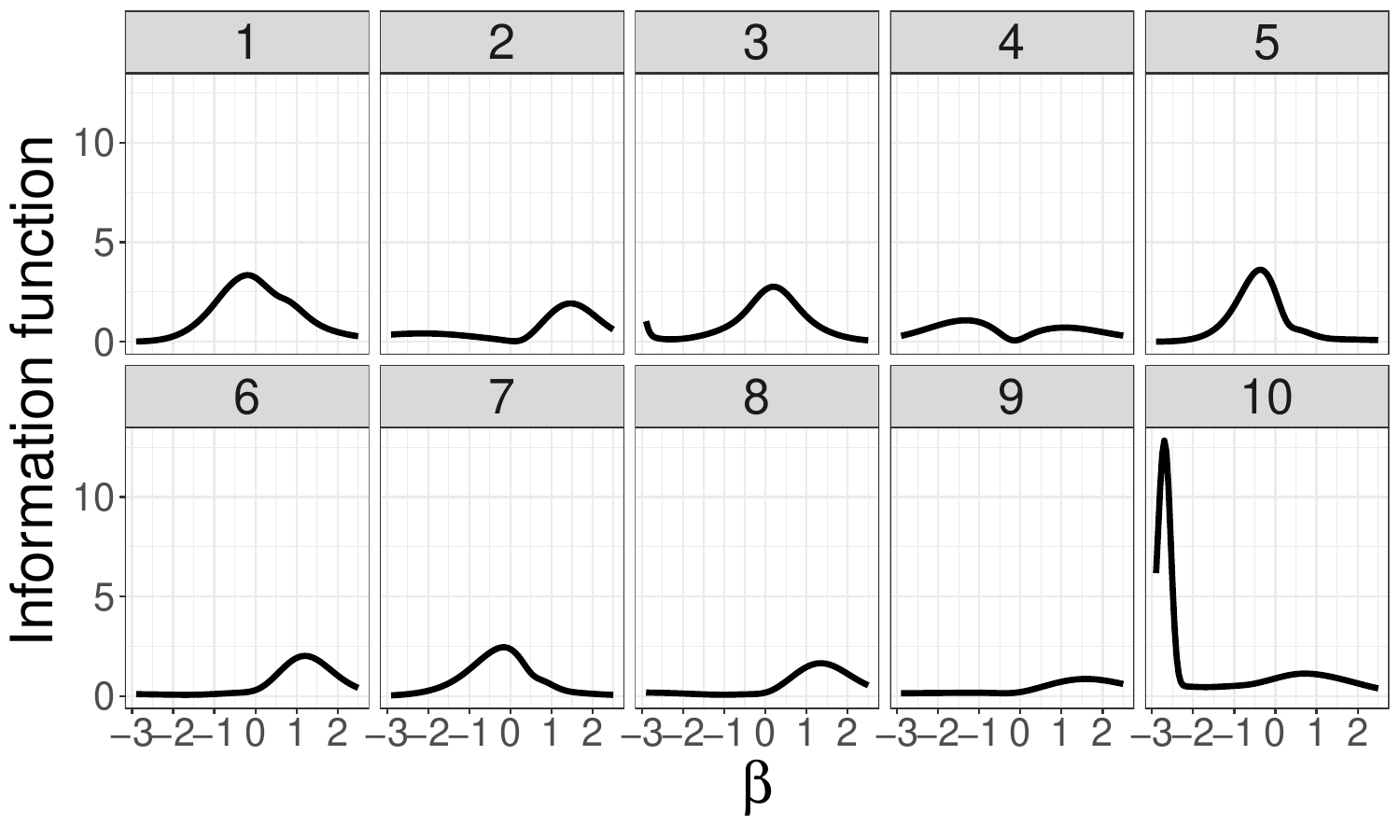}
    \caption{\editMark{Posterior mean of the information functions associated with each of the survey questions under OPUM.}}
    \label{fig:information_function}
\end{figure}

\editMark{Further insight can be gained by examining the information functions associated with the items, which are defined as  
\begin{align}
    I_j\left(\beta_i\right) &= \sum_{k = 0}^{K_j} \frac{1}{\theta_{i, j}(k)} \left(\frac{d }{d \beta_i}\theta_{i, j}(k)\right)^2 - \frac{d^2 }{d \beta_i^2} \theta_{i, j}(k), & j&=1,\ldots,10
    \label{eq:info_fct_prob}
\end{align}
(see the supplementary materials). Figure \ref{fig:information_function} shows the posterior mean of these information functions for each of the 10 survey questions under OPUM. Note that Question 1 and 2 provide information about different regions of the latent space:  while Question 2 seems to provide the least information about values of $\beta$ close to 0, Question 1 provide the most information about these.  On the other hand, most of the information that can be gleaned from Question 6 is about positive values of $\beta$, while Question 10 is particularly informative for individuals with extreme negative preferences.
}

\begin{figure}[!tb]
    \centering
    \begin{subfigure}[t]{0.38\textwidth}
        \includegraphics[width = \textwidth]{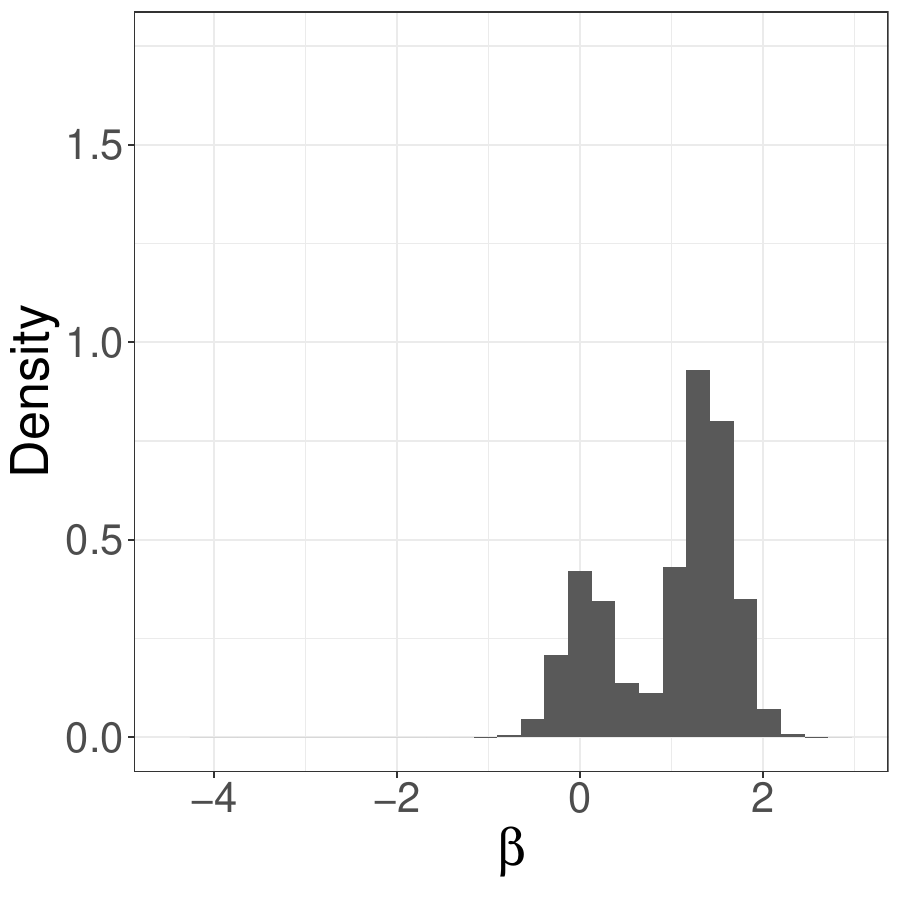}
        \caption{Individual 476}
    \end{subfigure}
    \begin{subfigure}[t]{0.38\textwidth}
        \includegraphics[width = \textwidth]{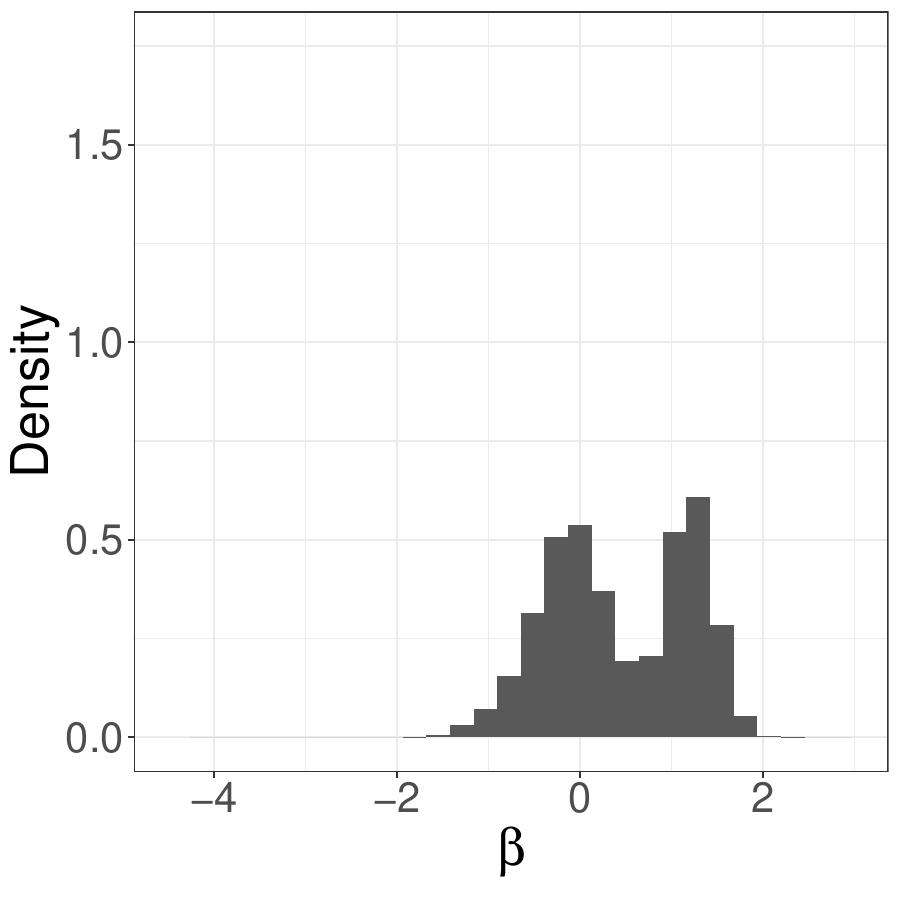}
        \caption{Individual 1192}
    \end{subfigure}\\
    \begin{subfigure}[t]{0.38\textwidth}
        \includegraphics[width = \textwidth]{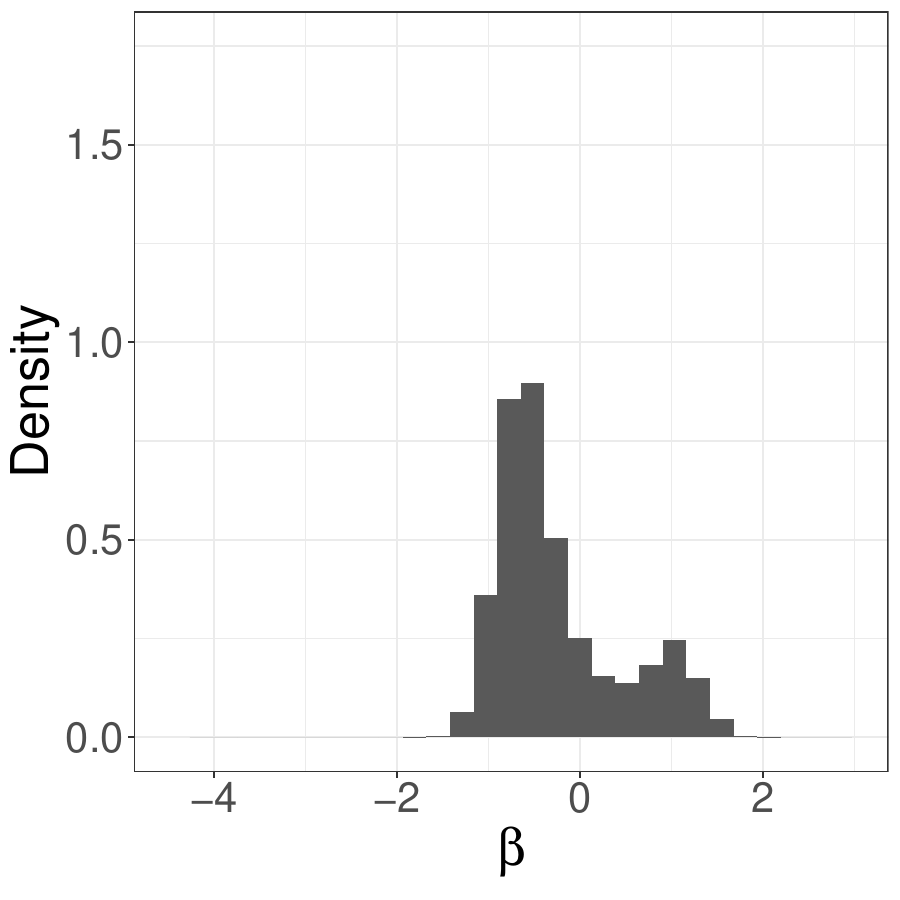}
        \caption{Individual 1273}
    \end{subfigure}
    \begin{subfigure}[t]{0.38\textwidth}
        \includegraphics[width = \textwidth]{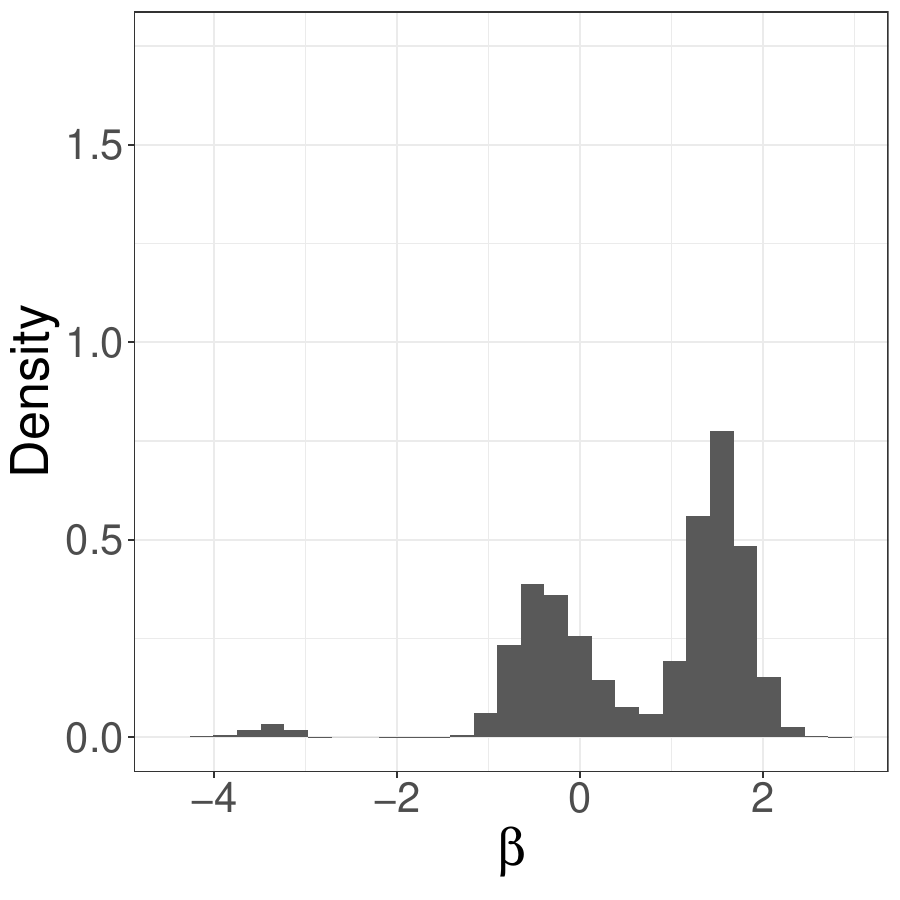}
        \caption{Individual 2193}
    \end{subfigure}
    \caption{Posterior distribution of the latent trait $\beta_i$ for selected individuals.}
    \label{fig:betaposterior}
\end{figure}

To conclude, we note that one underappreciated feature of models that allow for non-monotonic response functions is that the posterior distribution of the latent trait $\beta_i$ can be multi-modal for some individuals.  This multi-modality is not an artifact due to lack of identifiability or lack of mixing in the MCMC algorithms used to fit the models. 
Instead, it is a sign of real uncertainty and, potentially, of a poorly designed survey. To illustrate this, we present in Figure \ref{fig:betaposterior} the posterior distribution of $\beta_i$ for 4 selected individuals that exhibit multimodality.  Overall, about 8.1\% of the 2,621 respondents in this study show suchh multi-modal posteriors.\footnote{We use the function \texttt{is.multimodal()} in the \texttt{R} package \texttt{LaplacesDemon} with a minimum size of 0.1 to identify multimodal posterior distributions.}  Often, these were subjects who failed to express strong preferences in the items with clearly monotonic response functions (items 1 and 3). Meanwhile, 61\% of the 2621 respondents exhibit multi-modal posteriors under the BGGUM model.

\editMark{\subsection{Simulated data}

To further evaluate the performance of our model, we consider now a simulated dataset that aims to replicate the characteristics of the real data analyzed in Section \ref{sec:realdata}.  More specifically, we set $J=10$,  $I=2,621$, and $K_j=4$ for all $j$, and use the posterior means of the item-specific parameters $\mu_{j,-4} \ldots, \mu_{j,4}$ and $\alpha_{j,-4} \ldots, \alpha_{j,4}$ as the true values in the simulation (see Figure \ref{fig:true_resfunc_sim}). Similarly, the posterior mode of the latent traits $\beta_1, \ldots, \beta_I$ obtained in Section \ref{sec:realdata} were used as the true values of the latent traits.

\begin{figure}{!t}
    \centering
    \begin{subfigure}[t]{0.7\textwidth}
        \includegraphics[width = \textwidth]{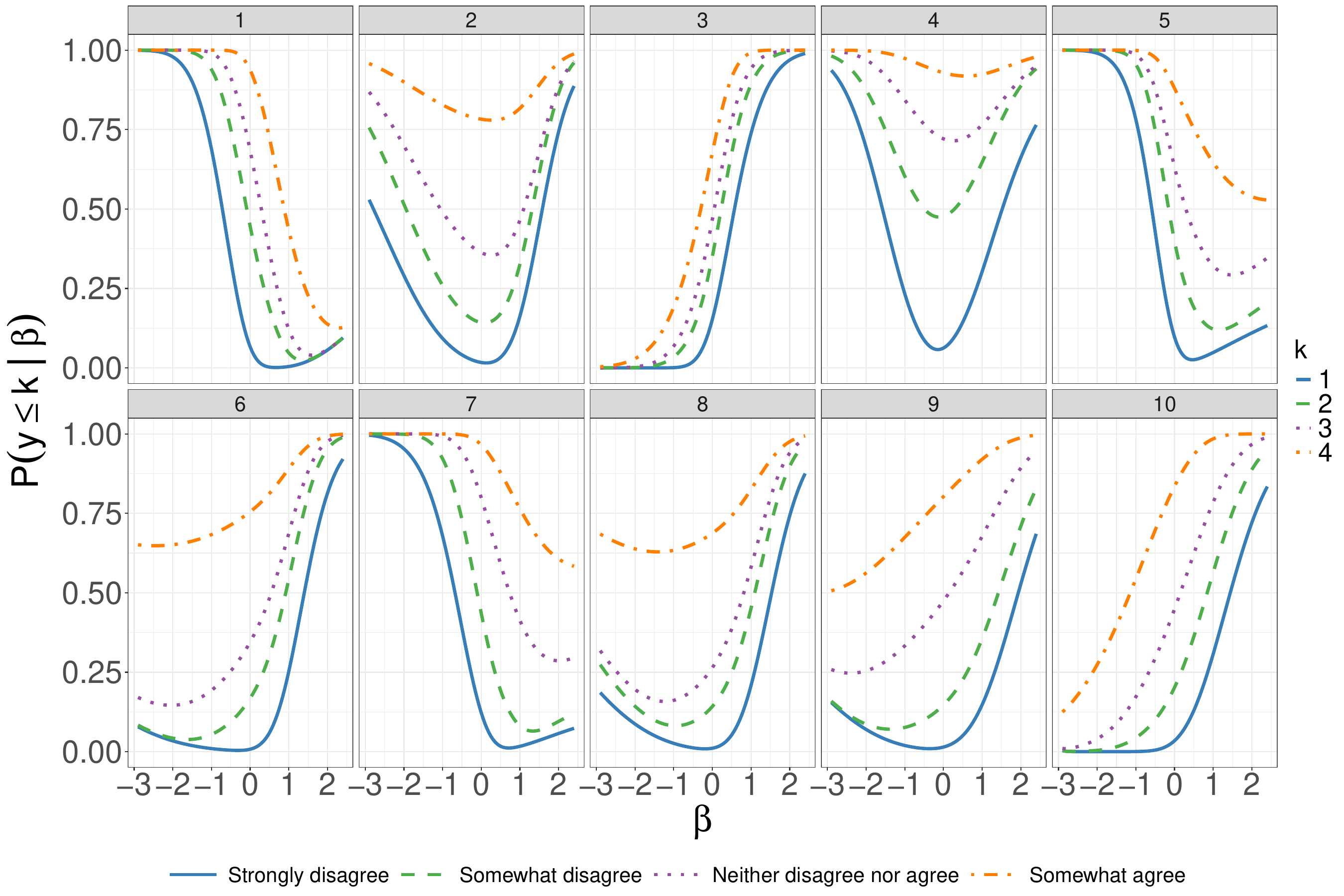}
        \centering
        \caption{True response functions used to generate the data in our simulation study.}\label{fig:true_resfunc_sim}
    \end{subfigure}\\
    \begin{subfigure}[t]{0.7\textwidth}
        \includegraphics[width = \textwidth]{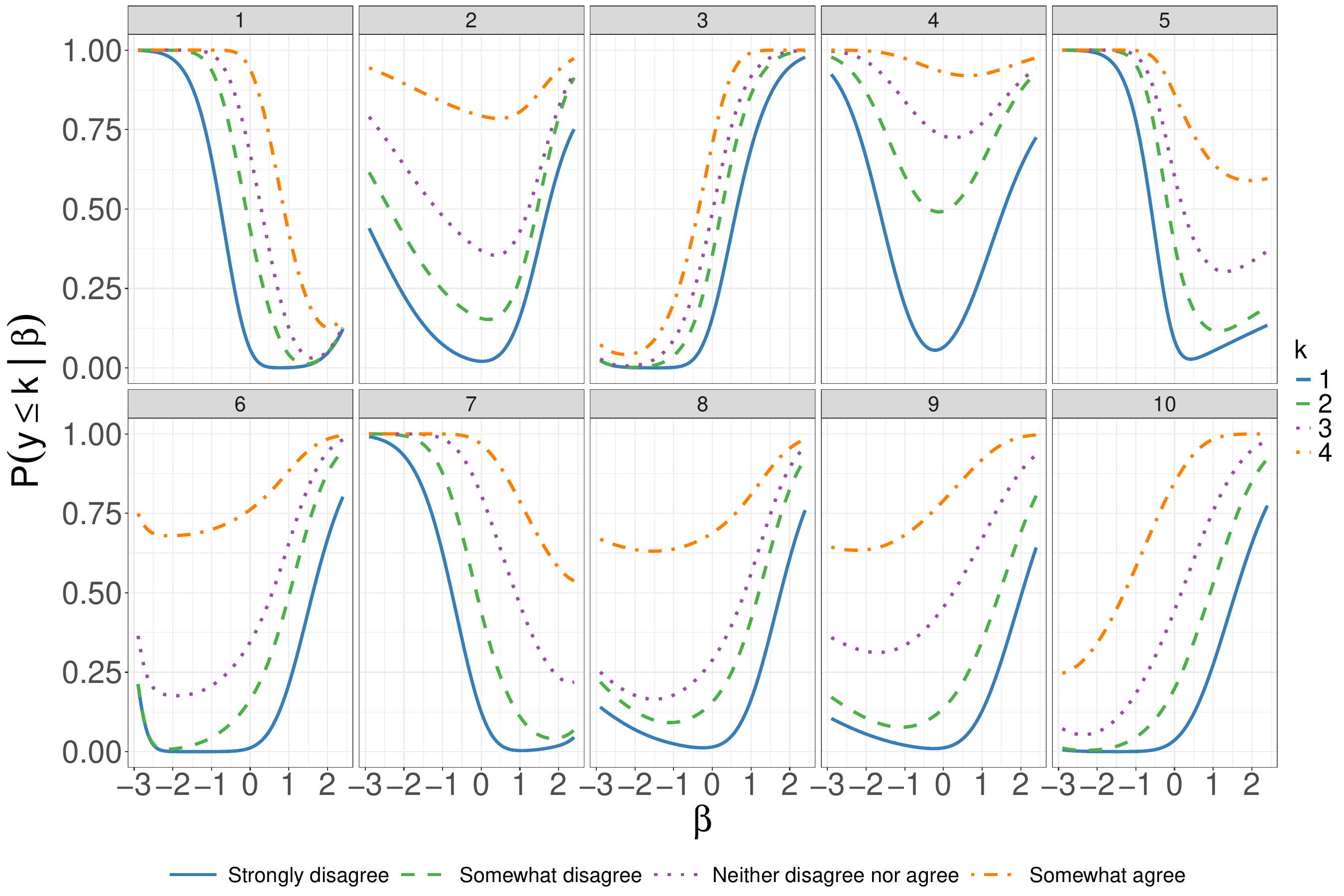}
        \caption{
        Posterior fitted mean response curves
        }\label{fig:est_resfunc_sim}
    \end{subfigure}
    \caption{Comparison of response curves from the simulated data based on simulated values and fitted values.}
    \label{fig:sim_response_curves_comp}
\end{figure}

We start by examining the coverage of 95\% posterior credible intervals. For the $\bfbeta_i$s and $\bfdelta_{j,k}$s, coverage is at least nominal: 95\% of the intervals for the $\delta_{j,k}$s and 95.7\% of the intervals for the $\beta_i$s include the true value. On the other hand, coverage for the $\alpha_{j,k}$s, at 91.25\%, is slightly below nominal. A closer look at the results indicates that these intervals tend to miss values of the $\alpha_{j,k}$s associated with items that have response functions that are monotonic over the observed range of latent traits. This is as it should be: the data available cannot provide any information about these parameters.  Additional intuition can be gained by comparing the posterior mean estimates of the response curves (Figure \ref{fig:est_resfunc_sim}) against their true shapes (Figure \ref{fig:true_resfunc_sim}).  While there are some small differences (e.g., in the estimate of $\Pr(y_{i,6} \le 0 \mid \beta_i)$ for very negative values of $\beta_i$), the estimated curves tend to be very similar to the true ones. This is consistent with the nominal coverage reported early.  Furthermore, note the $\alpha_{j,k}$ coefficients for which the credible intervals fail to contain the true values include $\alpha_{3,-4}$, $\alpha_{3,-3}$, $\alpha_{3,-2}$ and $\alpha_{3,-1}$, as well as $\alpha_{10,-4}$, $\alpha_{10,-3}$, $\alpha_{10,-2}$ and $\alpha_{10,-1}$.  Glancing at Figures \ref{fig:true_resfunc_sim} and
\ref{fig:est_resfunc_sim}, this is not surprising: these coefficients control the slopes of the unobserved left side of the response curves for items 3 and 10, respectively.}

\section{Discussion}\label{sec:discussion}

This paper introduced a novel ordinal probit unfolding model for polychotomous responses that can be fitted in a straightforward manner using a Markov chain Metropolis-Hastings algorithm with \editMark{minimal tuning}. The model is capable of accurately recovering the response functions of both monotonic and monotonic items and seems to provide better complexity-adjusted fit than available alternatives in real datasets. 

The results in Section 5 highlight are illuminating, as they highlight that the posterior distribution of latent traits for unfolding models like OPUM and BGGUM can be multimodal.  This issue seems to arise when the number of items in the survey is small and a large fraction of them have non-monotonic response functions (as is the case for the application in Section \ref{sec:results}). 
This makes intuitive sense:  non-monotonic items contain strong information about the absolute value of $\beta_i$ but limited information about its sign; information about the sign of $\beta_i$ is provided mainly by monotonic items. This problem might be particularly acute for BGGUM because of the symmetry constraint in their response functions might lead to further loss of information.  An extreme example of this can be seen in Figure \ref{fig:question_2_response_curve}.  In this case, the response function estimated by GGUM is (approximately) centered at 0 and, by construction, symmetric. This means that, under GGUM, Question 2 of the survey provides no information about the sign of the $\beta_i$s, as the value of the response function for $\beta_i = \Delta$ is essentially identical to that for $\beta_i = -\Delta$.  In contrast, the asymmetry in the response functions estimated by OPUM for Question 2 means that the model can extract at least a minimal amount of information about the sign of $\beta_i$ from the individual's response to this question.  

The previous observations have implications for survey design. They suggest that short instruments with mostly non-monotonic items might struggle to capture the correct directionality of the latent traits and should therefore be avoided or, at least, any latent traits estimated from them should treated with outmost care.  



Section \ref{sec:sym} discusses how OPUM (and, potentially, BGGUM as well) can be made agnostic to the positive/negative wording of a particular item.  In practice, this extension proved unnecessary in the application discussed in Section \ref{sec:results}, as our model did not identify any item in need of reverse coding. The pertinence of this result is confirmed by a careful reading of the 10 items in the survey, all of which can be expected to lead to either monotonic response functions, or non-monotonic ones in which the extremes of the latent scale would tend to strongly disagree.  Nonetheless, we think raising the visibility of this issue is important, as it has mostly been ignored in the literature, perhaps under the assumption that items have been carefully and appropriately designed with this limitation in mind.  Naively applying OPUM or BGGUM to reverse-coded items leads to flat estimates of their response functions, i.e., the inference that the item is not informative about the latent trait of interest.  At the very least, this means a loss in statistical efficiency in the estimation of the latent trait.  In extreme circumstances, this can also lead to patently inaccurate estimates of the latent traits.

There are a few interesting directions to expand this work. \editMark{An obvious extension is to consider multivariate latent traits. One way to accomplish this is to first rewrite $\eta_{i,j,k} = \alpha_{j,k}\beta_i - \mu^{*}_{j,k}$, where $\mu^{*}_{j,k} = \alpha_{j,k}\mu_{j,k}$ in Section \ref{sec:model}, and then replace the product $\alpha_{j,k}\beta_i$ with a linear combination $\bfalpha^T_{j,k}\bfbeta_i$ where $\bfalpha_{j,k}, \bfbeta_i \in \real^{D}$.  Inference on $D$ could be accomplished using the same WAIC criteria discussed in Section \ref{sec:results}.} Relaxing the assumption of independence in the item parameters is another interesting avenue to explore. This might be particularly important in situations where individuals might anchor the responses to later survey questions on the responses provided in earlier ones.  In such cases, the fact that the anchor might vary across subjects means that some of the independence assumptions we relied on to elicit our priors are not appropriate.  \editMark{Finally, we could consider other utility functions, such as those based on Manhattan (L1) distances (e.g., see \citealp{Shin03042025}), Guassian utlity functions (e.g., see \citealp{carroll2013structure}), or other such functions (e.g. see \citet{de1983mixed}).  More generally, we could use Bayesian nonparametric techniques to learn the shape of the utility functions. These extensions, however, would likely involve the use of more  cumbersome computational approaches that those described in this paper.  Furthermore, evidence from the use of these alternative utility functions in standard spatial models suggests that they tend to have an impact mostly in settings with multivariate latent spaces (e.g., see \citealp{Shin03042025}, \citealp{grynaviski2006specification}, and \citealp{carroll2013structure}).}



\printbibliography


\setcounter{section}{0}
\section*{\Large Supplementary Material}
\section{Properties of the GGUM response functions}

\subsection{Minimum}

The fact that the minimum of the response function for a BGGUM corresponds to $\beta_i = \mu_j$ is \cite{duck2022ends}, who show that this is the point where the probability of selecting response folds on itself. Indeed, we can easily check that 
$$
\left. \frac{\partial }{\partial \beta_i} \Pr(y_{i,j} \le k \mid \beta_i) \right|_{\beta_i = \mu_j} = 0
$$
and that, for all $k \in \{ 0, \ldots, K_j \}$, $\frac{\partial }{\partial \beta_i} \Pr(y_{i,j} \le k \mid \beta_i) < 0$ for $\beta_i < \mu_j$ and $\frac{\partial }{\partial \beta_i} \Pr(y_{i,j} \le k \mid \beta_i) > 0$ for $\beta_i > \mu_j$.

\subsection{Symmetry}
It suffices to show that $\Pr(y_{i,j} = k \mid \beta_i = \mu_j + \Delta) = \Pr(y_{i,j} = k \mid \beta_i = \mu_j - \Delta)$ and thus that $\Pr(y_{i,j} \le k \mid \beta_i = \mu_j + \Delta) = \Pr(y_{i,j} \le k \mid \beta_i = \mu_j - \Delta)$. Without loss of generality, we assume that $\mu_j = 0$. 

Note that because of the constraints on $\v{\tau}$, we have that for all k = $0, 1, \ldots, K_j$,
\[
 \sum_{l=0}^{k} \tau_{j,l} = \sum_{l=0}^{2K_j+1 - k} \tau_{j,l}.
\]
As a result, conditioned on $\alpha_j, \Delta$, we have that
\begin{align*}
    & \Pr(y_{i,j} = k \mid \beta_i = \Delta)\\ 
    &= \frac{\exp\left\{ \alpha_j \left[  k \Delta - \sum_{l=0}^{k} \tau_{j,l} \right] \right\} + \exp\left\{ \alpha_j \left[  \left(2K_j+1 - k\right) \Delta - \sum_{l=0}^{k} \tau_{j,l} \right] \right\}}
    {\sum_{k'=0}^{K} \left\{ \exp\left\{ \alpha_j \left[  k \Delta - \sum_{l=0}^{k'} \tau_{j,l} \right] \right\} + \exp\left\{ \alpha_j \left[  \left(2K_j+1 - k\right) \Delta - \sum_{l=0}^{k'} \tau_{j,l} \right] \right\}\right\}}\\
    &=  \frac{\exp\left(- \alpha_j\sum_{l=0}^{k} \tau_{j,l}\right) \left(\exp\left\{ \alpha_j (2K_j + 1 - k) (-\Delta) \right\} + \exp\left\{ \alpha_j k (-\Delta)\right\} \right)}
    {\sum_{k'=0}^{K} \left\{ \exp\left\{ \alpha_j \left[ k (-\Delta) - \sum_{l=0}^{k'} \tau_{j,l} \right] \right\} + \exp\left\{ \alpha_j \left[  \left(2K_j+1 - k\right) (-\Delta) - \sum_{l=0}^{k'} \tau_{j,l} \right] \right\}\right\}}\\
    &= \Pr(y_{i,j} = k \mid \beta_i = -\Delta).
\end{align*}

\subsection{Limit as $| \beta_i | \to \infty$}

Note that, for $k' > k$ and as $\beta_i \rightarrow \infty$,
\begin{multline*}
\frac{\exp\left\{ \alpha_j \left[  \left(2K_j+1 - k\right) (\beta_i - \mu_j) - \sum_{l=0}^{2K_j+1 - k} \tau_{j,l} \right] \right\}}{\exp\left\{ \alpha_j \left[  \left(2K_j+1 - k'\right) (\beta_i - \mu_j) - \sum_{l=0}^{2K_j+1 - k'} \tau_{j,l} \right] \right\}} \\
= \exp\left\{ \alpha_j \left[  \left(k' - k\right) (\beta_i - \mu_j) - \sum_{l=2K_j+1 - k'}^{2K_j+1 - k} \tau_{j,l} \right] \right\} \rightarrow \infty.
\end{multline*}

As a result, the subjective probability associated with the first category grows the fastest as $\beta_i \rightarrow \infty$ and will dominate the other subjective probability. Hence, $p(y = 0 \mid \beta) \rightarrow 1$ as $\beta \rightarrow \infty$ so
$p(y \leq k \mid \beta_i) \rightarrow 1$ for all $k \in 0, 1, \ldots, K_j$.  The limit as $\beta_i \to -\infty$ is the same because of the symmetry of the response function.

\section{Properties of OPUM}

\subsection{Computing $\theta_{i,j}(k)$ under OPUM}

Because the errors are normal and based on (3.2), we can explicitly compute the probability as the following using the $\eta_{i, j, k}$ (and thus the $\bfalpha_j$'s and $\bfmu_j$'s):

\begin{align*}
    P(y_{i, j} \mid \beta_i, \alpha_{j}, \delta_j) = \theta_{i,j}^{OPUM}(k) = \begin{cases}
      \nu_{i,j}(0)  & k=0 , \\
      \nu_{i,j}(-k) + \nu_{i,j}(k)  & k \ge 1,
    \end{cases}
\end{align*}
where
\begin{multline*}
  \nu_{i,j}(k) = \\
  \int_{\eta_{i, -K_j} - \eta_{i, k}}^{\infty} \int_{\eta_{i, -K_j + 1} - \eta_{i, k}}^{\infty} \cdots \int_{\eta_{i, k - 1} - \eta_{i, k}}^{\infty} \int_{\eta_{i, k + 1} - \eta_{i, k}}^{\infty} \cdots \int_{\eta_{i, K_j} - \eta_{i, k}}^{\infty} \phi_{2K_j}\left(
    \v{z}^{-k} \mid \v{0}_{2K_j}, \bfSigma_j \right) d \v{z},
\end{multline*}
$\v{z}^{-k} \in \mathbbm{R}^{2K_j}$ is a vector such that for $k' \in \{ -K_j, -K_j + 1, \ldots, K_j \}$, $k' \neq k$, $z^{-k}_{i, j, k'} = z_{i, j, k} - z_{i, j, k'}$,
$\phi_{m} \left(\cdot \mid \v{d}, \v{D} \right)$ denotes the density of the $m$-variate normal distribution with mean vector $\v{d}$ and covariance matrix $\v{D}$, $\mathbf{0}_{d}$ is a vector of zeros of dimension $d$, and $\bfSigma_j$ is the $2K_j \times 2K_j$ matrix with entries $[\bfSigma_j]_{l, l} = 2$ when $l = l'$ and $[\bfSigma_j]_{l, l'} = 1$ for $l \ne l'$.

\subsection{Identifiability}
We follow the logic in \cite{lei2025logit}. For our model, the posterior distribution is given by 
\begin{multline*}
\textrm{Pr}(\{\beta_i\}, \{ \mu_{j,k} \}, \{ \alpha_j \}, \{ \zeta_j \}  \mid \mathbf{Y}) \propto \\
\textrm{Pr}\left(\bfY \mid \{\beta_i\}, \{ \mu_{j,k} \}, \{ \alpha_j \}, \{ \zeta_j \} \right) \prod_{i=1}^{I} \textrm{Pr}(\beta_i)  \prod_{j=1}^{J}  \textrm{Pr}(\{ \mu_{j,k} \}, \{ \alpha_j \}) \textrm{Pr}(\zeta_j).
\end{multline*}

Here, the likelihood is given by
\begin{multline*}
    \textrm{Pr}\left(\bfY \mid \{\beta_i\}, \{ \mu_{j,k} \}, \{ \alpha_j \}, \{ \zeta_j \} \right) = \\
    \prod_{j=1}^{J} \prod_{k=0}^{K_j} \left\{ \prod_{i=1}^{I} \left\{ \theta_{i,j}(k) \right\}^{\indFct{y_{i,j}=k} \indFct{\zeta_{j}=0}}  \left\{\theta_{i,j}(K_j-k) \right\}^{\indFct{y_{i,j}=k} \indFct{\zeta_{j}=1}}
    \right\},
\end{multline*}
where
\begin{align*}
    \theta_{i,j}^{OPUM}(k) = \begin{cases}
      \nu_{i,j}(0)  & k=0 , \\
      \nu_{i,j}(-k) + \nu_{i,j}(k)  & k \ge 1.
    \end{cases}
\end{align*}

Note that we define
\begin{align*}
  \nu_{i,j}(k) = 
  \int_{\eta_{i, -K_j} - \eta_{i, k}}^{\infty} \int_{\eta_{i, -K_j + 1} - \eta_{i, k}}^{\infty} \cdots \int_{\eta_{i, k - 1} - \eta_{i, k}}^{\infty} \int_{\eta_{i, k + 1} - \eta_{i, k}}^{\infty} \cdots \int_{\eta_{i, K_j} - \eta_{i, k}}^{\infty} \phi_{2K_j}\left(
    \v{z}^{-k} \mid \v{0}_{2K_j}, \bfSigma_j \right) d \v{z},
\end{align*}
$\v{z}^{-k} \in \mathbbm{R}^{2K_j}$ is a vector such that for $k' \in -K_j, -K_j + 1, \ldots, K_j$, $k' \neq k$, $z^{-k}_{i, j, k'} = z_{i, j, k} - z_{i, j, k'}$,
$\phi_{m} \left(\cdot \mid \v{d}, \v{D} \right)$ denotes the density of the $m$-variate normal distribution with mean vector $\v{d}$ and covariance matrix $\v{D}$, $\mathbf{0}_{d}$ is a vector of zeros of dimension $d$, and $\bfSigma_j$ is the $2K_j \times 2K_j$ matrix with entries $[\bfSigma_j]_{l, l} = 2$ when $l = l'$ and $[\bfSigma_j]_{l, l'} = 1$ for $l \ne l'$. 

As $\eta_{i,j,k} = \alpha_{j,k} \left( \beta_i - \mu_{j,k} \right)$, we can re-parametrize in the following manner for some real constants $a > 0$ and $b$ and $k = -K_j, -K_j + 1, \ldots, -K_j$:
\begin{itemize}
    \item $\alpha'_{j,k} = \frac{1}{a} \alpha_{j,k}$ 
    \item $\mu'_{j,k} = a (\mu_{j,k} + b)$ 
    \item $\beta'_{i} = a (\beta_{i} + b)$ 
\end{itemize}
Then, we have that for all $k$,
\begin{align*}
    \eta'_{i,j,k} &= \alpha'_{j, k} (\beta'_i - \delta'_{j, k})\\ 
    &= \frac{1}{a}\alpha_{j, k} \left\{ a (\beta_i + b) - a (\delta_{j, k} + b) \right\}\\
    &= \alpha_{j, k} \left\{ (\beta_i + b) - (\delta_{j, k} + b) \right\} \\
    &= \alpha_{j, k} (\beta_i - \delta_{j, k})\\
    &= \eta_{i,j,k}.
\end{align*}
As a result,
\begin{align*}
    \textrm{Pr}\left(\bfY \mid \{\beta_i\}, \{ \mu_{j,k} \}, \{ \alpha_j \}, \{ \zeta_j \} \right) &= 
    \textrm{Pr}\left(\bfY \mid \{\beta'_i\}, \{ \mu'_{j,k} \}, \{ \alpha'_j \}, \{ \zeta_j \} \right).
\end{align*}

This demonstrates that under the likelihood alone, the parameters are not identifiable under shifts and rescalings.  However, note that $\beta_i'$'s prior distribution is a normal one with mean $ab$ and variance $a^2$.  Thus, $\textrm{Pr}(\beta_i') \ne \textrm{Pr}(\beta_i)$, leading to different posterior probabilities. In other words, under shifts and rescaling, the model is identified only through the choice of prior distribution and is thus weakly identifiable.

On the other hand, consider the following re-parametrization for $k = -K_j, -K_j + 1, \ldots, -K_j$:
\begin{itemize}
    \item $\alpha''_{j,k} = -\alpha_{j,-k}$ 
    \item $\mu''_{j,k} = -\mu_{j,-k}$ 
    \item $\beta''_{i} = -\beta_{i}$  
\end{itemize}
Then, we have that for all $k$,
\begin{align*}
    \eta''_{i,j,k} &= \alpha''_{j, k} (\beta''_i - \delta''_{j, k})\\ 
    &= -\alpha_{j, -k} \left\{ -\beta_i - -\delta_{j, -k} \right\}\\
    &= \alpha_{j, -k} (\beta_i - \delta_{j, -k})\\
    &= \eta_{i,j,-k}.
\end{align*}
If we define $\nu''_{i,j}(k)$ via $\eta''_{i,j,k}$, it is clear that
\[
\nu''_{i,j}(k) = \nu_{i,j}(-k).
\]
Still, as $\theta_{i,j}^{OPUM}(k) = \nu_{i,j}(k) + \nu_{i,j}(-k)$ for $k \neq 0$,
\begin{align*}
    \textrm{Pr}\left(\bfY \mid \{\beta_i\}, \{ \mu_{j,k} \}, \{ \alpha_j \}, \{ \zeta_j \} \right) &= 
    \textrm{Pr}\left(\bfY \mid \{\beta''_i\}, \{ \mu''_{j,k} \}, \{ \alpha''_j \}, \{ \zeta_j \} \right).
\end{align*}
Here, as $p(\beta_i') = p(\beta_i)$ and our prior for the $\alpha_j$'s and $\delta_j$'s are centered around 0, the posterior probability will be the same. As a result, the prior does not address this type of unidentifability. However, there is a straightforward approach to address this issue. We will fix the sign of a respondent so that the scales between all three models are the same. To accomplish this, we will take advantage of the same reparametrization that led to this unidentifiability.

\section{Gibbs sampler}

Our Gibbs sampler is as following:
\begin{enumerate}
    \item Sample $y^*_{i, j, m} \mid \cdots$ for all $i = 1, 2, \ldots, I$, $j = 1, 2, \ldots, J$, and $m = -(K - 1), -(K - 2), \ldots, K - 1$ from its full posterior conditional distribution:
    \begin{align*}
        & y^*_{i, j, m} \mid \cdots \sim\\
        & \quad
        \begin{cases}
            \textrm{N}(\cdot \mid \eta_{i, j, m}, 1)\indFct{y^*_{i, j, m} < \max(y^*_{i, j, -(K - y_{i, j})}, y^*_{i, j, (K - y_{i, j})})} & y_{i, j} \notin \{-(K - m), K - m\},\\
            \textrm{N}(\cdot \mid \eta_{i, j, m}, 1)\indFct{y^*_{i, j, m} > \max_{m' \neq m}(y^*_{i, j, m'})} & y_{i, j} \in \{-(K - m), K - m\};\\
            & y^*_{i, j, -m} < \max_{m' \neq m, -m}(y^*_{i, j, m'}),\\
            \textrm{N}(\cdot \mid \eta_{i, j, m}, 1) & \textrm{o.w.}\\
        \end{cases}       
    \end{align*}
    Note that if $m = 0$, this simplifies to 
    \begin{align*}
        y^*_{i, j, m} \mid \cdots &\sim 
        \begin{cases}
            \textrm{N}(\cdot \mid 0, 1)\indFct{y^*_{i, j, 0} < \max(y^*_{i, j, -(K - y_{i, j})}, y^*_{i, j, (K - y_{i, j})})} & y_{i, j} \neq 0,\\
            \textrm{N}(\cdot \mid 0, 1) \indFct{y^*_{i, j, 0} > \max_{m' \neq 0}(y^*_{i, j, m'})} & y_{i, j} = 0.\\
        \end{cases}       
    \end{align*}
    \item Sample $\beta_i \mid \cdots$ for all $i = 1, 2, \ldots, I$ from its full posterior conditional distribution, $\textrm{N}(\cdot \mid \mu_{\beta_i}, \bfSigma^2_{\beta_i})$, where
    \begin{align*}
    \sigma^2_{\beta_i} = \frac{1}{1 + \sum_{j =1}^J (\bfalpha_j^T \bfalpha_j)}, & &\mu_{\beta_i} = \frac{\sum_{j = 1}^J \v{\alpha_{j}}^T(\v{y^*_{i, j}} + \bfD_{\bfalpha_j} \bfmu_j)}{1 + \sum_{j =1}^J (\bfalpha_j^T \bfalpha_j)}.
    \end{align*}
    Here, $\bfD_{\bfalpha_j} = \textrm{diag}(\bfalpha_j)$.
    \item Sample $\bfalpha_j \mid \cdots$ for all $j = 1, 2, \ldots, J$ from its full posterior conditional distribution, a truncated multivariate normal distribution of dimension $2(K - 1)$:
    \begin{align*}
        \textrm{N}_{2(K - 1)} & \left(\cdot \bigg|
            \begin{pmatrix}
                \v{\mu}_{\bfalpha_j, 1}\\ 
                \v{\mu}_{\bfalpha_j, 2}
            \end{pmatrix}
            , 
            \begin{pmatrix}
                \bfSigma_{\bfalpha_j, 1} & 0\\
                0 & \bfSigma_{\bfalpha_j, 2}
            \end{pmatrix}\right)\\
            & \indFct{\alpha_{j, -(K - 1)} < \alpha_{j, -(K - 2)} \ldots < \alpha_{j, - 1} < 0 < \alpha_{j, 1}< \ldots < \alpha_{j, K - 1}},
    \end{align*}
    \item For every 50 iterations, propose    
    $y'_{i, j} = K - k + 1$ and $\bfalpha'_j$ and $\bfmu'_j$ from the prior defined in (6) for each question $j = 1, 2, \ldots, J$ and all individuals $i$ who answer question $j$. Accept the proposal with probability
    \[
    \min\left(1, \prod_{i = 1}^I \frac{p(y'_{i, j} \mid \beta_i, \bfalpha'_j, \bfmu'_j)}{p(y_{i, j} \mid \beta_i, \bfalpha_j, \bfmu_j)}  \right).
    \]
    \item For every 50 iterations, propose $\beta'_i \sim \textrm{N}(\cdot \mid -\beta_i, 0.75)$. Accept this proposal with probability
    \[
    \min\left(1, \prod_{j = 1}^J \frac{p(y_{i, j} \mid \beta'_i, \bfalpha_j, \bfmu_j)\textrm{N}(\beta'_i \mid 0, 1)}{p(y_{i, j} \mid \beta_i, \bfalpha_j, \bfmu_j)\textrm{N}(\beta_i \mid 0, 1)}  \right).
    \]
            
        where $\v{\mu}_{\bfalpha_j, 1}, \v{\mu}_{\bfalpha_j, 2}$ are the vectors and $\bfSigma_{\bfalpha_j, 1}, \bfSigma_{\bfalpha_j, 2}$ are the matrices such that
        \begin{align*}
        \bfSigma_{\bfalpha_j, 1} = \left(\bfXi_1^{-1} + \sum_{i = 1}^I \bfD^2_{\beta_i \v{\mathbbm{1}}_{K - 1} - \bfmu^{-}_j}\right)^{-1}, & &
        \v{\mu}_{\bfalpha_j, 1} = \bfSigma_{\bfalpha_j, 1}\left(\sum_{i = 1}^I \bfD_{\beta_i \v{\mathbbm{1}}_{K - 1} - \bfmu^{-}_j} \bfy^{-}_{i, j}\right), 
        \\
        \bfSigma_{\bfalpha_j, 2} = \left(\bfXi_2^{-1} + \sum_{i = 1}^I \bfD^2_{\beta_i \v{\mathbbm{1}}_{K - 1} - \bfmu^{+}_j}\right)^{-1}, & &
        \v{\mu}_{\bfalpha_j, 2} = \bfSigma_{\bfalpha_j, 2}\left(\sum_{i = 1}^I \bfD_{\beta_i \v{\mathbbm{1}}_{K 
        - 1} - \bfmu^{+}_j} \bfy^{+}_{i, j}\right).
    \end{align*}
    Note that 
    $\bfD_{\beta_i \v{\mathbbm{1}}_{K 
        - 1} - \bfmu^{-}_j} = \textrm{diag}(\beta_i \v{\mathbbm{1}}_{K 
        - 1} - \bfmu^{-}_j)$, 
    $\bfD_{\beta_i \v{\mathbbm{1}}_{K 
        - 1} - \bfmu^{+}_j} = \textrm{diag}(\beta_i \v{\mathbbm{1}}_{K 
        - 1} - \bfmu^{+}_j)$,
    $\bfmu^{-}_{j} = \{\mu_{j, m}\}_{m = -(K - 1)}^{-1}$, $\bfmu^{+}_{j} = \{\mu_{j, m}\}_{m = 1}^{K - 1}$, $\bfy^{-}_{i, j} = \{y_{i, j, m}\}_{m = -(K - 1)}^{-1}$, $\bfy^{+}_{i, j} = \{y{i, j, m}\}_{m = 1}^{K - 1}$,
    and $\mathbbm{1}_{L}$ denotes a vector of ones of length L. 
    
        
    \item Sample $\bfmu_j \mid \cdots$ for all $j = 1, 2, \ldots, J$ from its full posterior conditional distribution, $\textrm{N}_{2(K - 1)}(\cdot \mid \v{\mu}_{\bfmu_j}, \bfSigma_{\bfmu_j})$, such that
    \begin{align*}
    \bfSigma_{\bfmu_j} = (\bfOmega^{-1} + I \bfD^2_{\bfalpha_j})^{-1}, && \v{\mu}_{\bfmu_j} = \bfSigma_{\bfmu_j}\left( \bfOmega^{-1} \bftheta + \sum_{i = 1}^I \bfD_{\bfalpha_j}(\beta_i \bfalpha_j - \v{y^*_{i,j}})\right).
    \end{align*}
\end{enumerate}

\section{Convergence diagnostics}

Below, we share the converge diagnostics on the $\hat{R}$ and effective sample size for the parameters. Notice that the order of the responses were never flipped and that $\psi_0$ represents the option associated with the highest level of agreement.

\begin{table}[ht]
\centering
\begin{minipage}{0.48\textwidth}
\centering
\begin{tabular}{lccc}
\hline
\textbf{Parameter} & \textbf{Median} & \textbf{Min} & \textbf{Max} \\
\hline
&\\
$\beta_i$   & 1.00 & 1.00 & 1.89  \\
& \\
$\alpha_{j, k}$  & 1.04  & 1.00  & 3.04 \\
& \\
$\delta_{j, k}$  & 1.02 & 1.00 & 2.93 \\
& \\
\hline
\end{tabular}
\caption{$\hat{R}$ for parameters}
\label{tab:params_summary}
\end{minipage}
\hfill
\begin{minipage}{0.48\textwidth}
\centering
\includegraphics[width=0.8\textwidth]{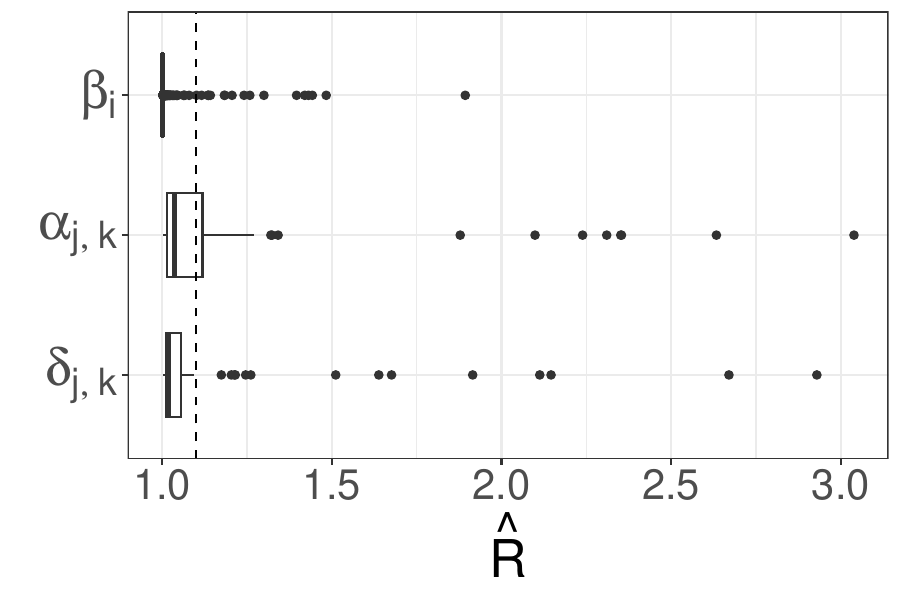}
\captionof{figure}{$\hat{R}$ boxplot}
\end{minipage}
\end{table}

\begin{table}[ht]
\centering
\begin{minipage}{0.48\textwidth}
\centering
\begin{tabular}{lccc}
\hline
\textbf{Parameter} & \textbf{Median} & \textbf{Min} & \textbf{Max} \\
\hline
&\\
$\beta_i$   & 6272.71 & 2.26 & 11738.29  \\
& \\
$\alpha_{j, k}$  & 96.65  & 1.31  & 1878.30 \\
& \\
$\delta_{j, k}$  & 173.67 & 1.39 & 2714.42 \\
& \\
\hline
\end{tabular}
\caption{Effective sample size (bulk) for parameters}
\label{tab:params_summary}
\end{minipage}
\hfill
\begin{minipage}{0.48\textwidth}
\centering
\includegraphics[width=0.8\textwidth]{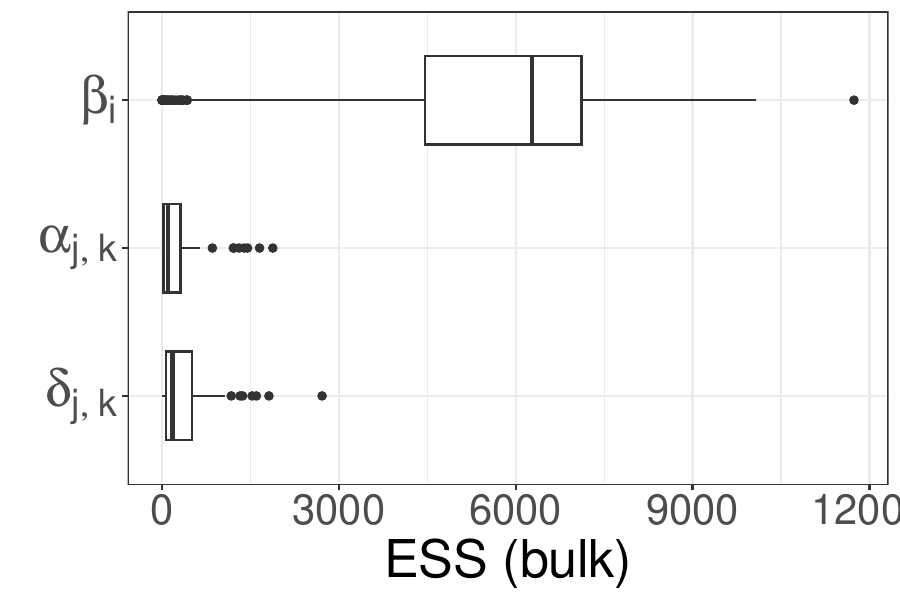}
\captionof{figure}{Effective sample size (bulk) boxplots}
\end{minipage}
\end{table}

\begin{table}[ht]
\centering
\begin{minipage}{0.48\textwidth}
\centering
\begin{tabular}{lccc}
\hline
\textbf{Parameter} & \textbf{Median} & \textbf{Min} & \textbf{Max} \\
\hline
&\\
$\beta_i$   & 11645.67  & 11.49 & 20582.32  \\
& \\
$\alpha_{j, k}$  & 327.49  & 11.31  & 5205.04   \\
& \\
$\delta_{j, k}$  & 451.15 & 10.51 & 10283.46 \\
& \\
\hline
\end{tabular}
\caption{Effective sample size (tail) for parameters}
\label{tab:params_summary}
\end{minipage}
\hfill
\begin{minipage}{0.48\textwidth}
\centering
\includegraphics[width=0.8\textwidth]{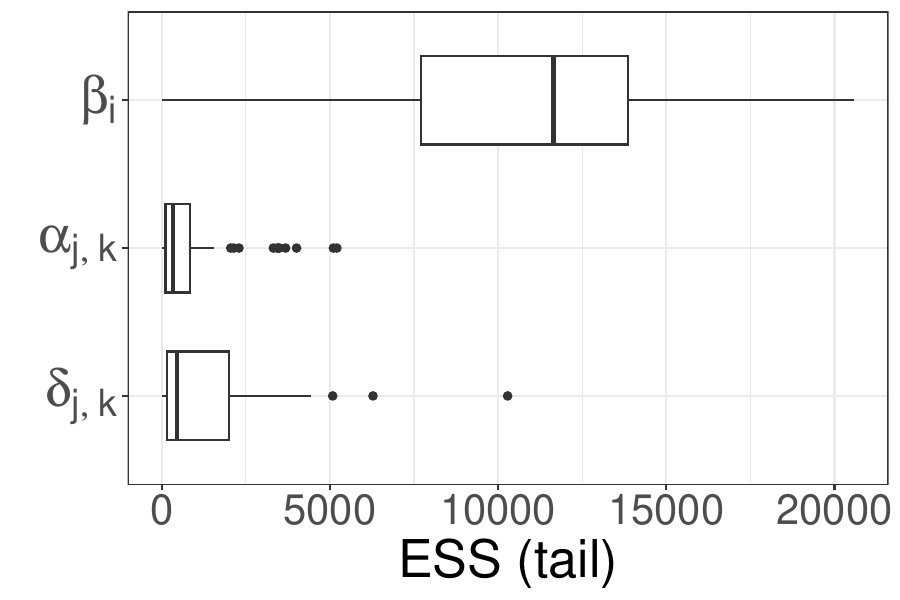}
\captionof{figure}{Effective sample size (tail) boxplots}
\end{minipage}
\end{table}

\begin{table}[]
    \centering
    \begin{tabular}{ccccc}
\hline
\textbf{Chain} & \textbf{Flip number} & \textbf{Median} & \textbf{Min} & \textbf{Max} \\
\hline
&\\
1 & 546   & 0.028   & 0.00 & 0.223   \\
& \\
2 & 236 & 0.131   & 0.00  & 0.549    \\
& \\
3 & 221  & 0.142  & 0.00 & 0.558  \\
& \\
4 & 209 & 0.178 & 0.00 & 0.585 \\
& \\
\hline
\end{tabular}
\caption{Acceptance probability}
\end{table}

\begin{figure}[!tb]
    \centering
    \begin{subfigure}[t]{0.3\textwidth}
        \includegraphics[width = \textwidth]{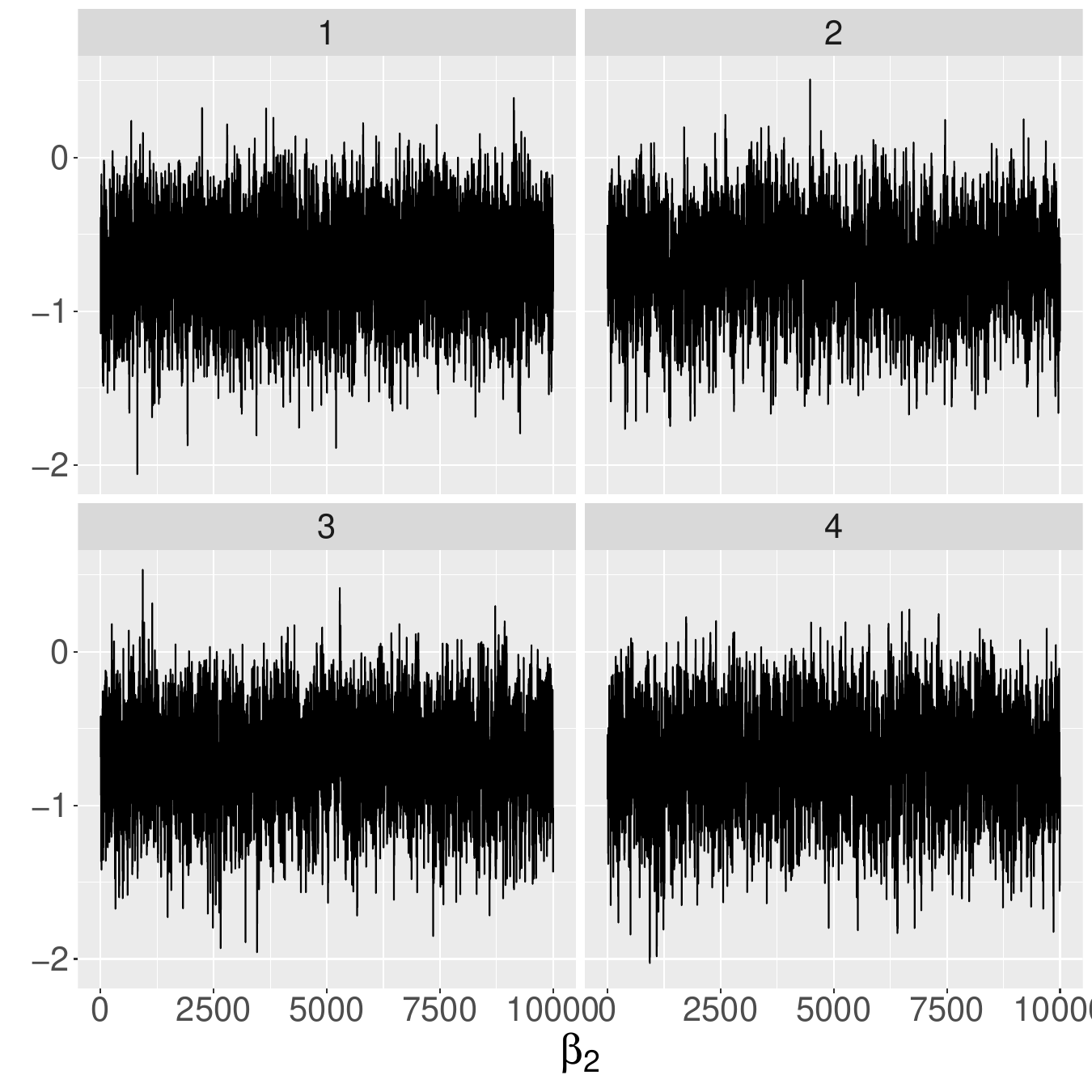}
        \caption{Individual 2}
    \end{subfigure}
    \begin{subfigure}[t]{0.3\textwidth}
        \includegraphics[width = \textwidth]{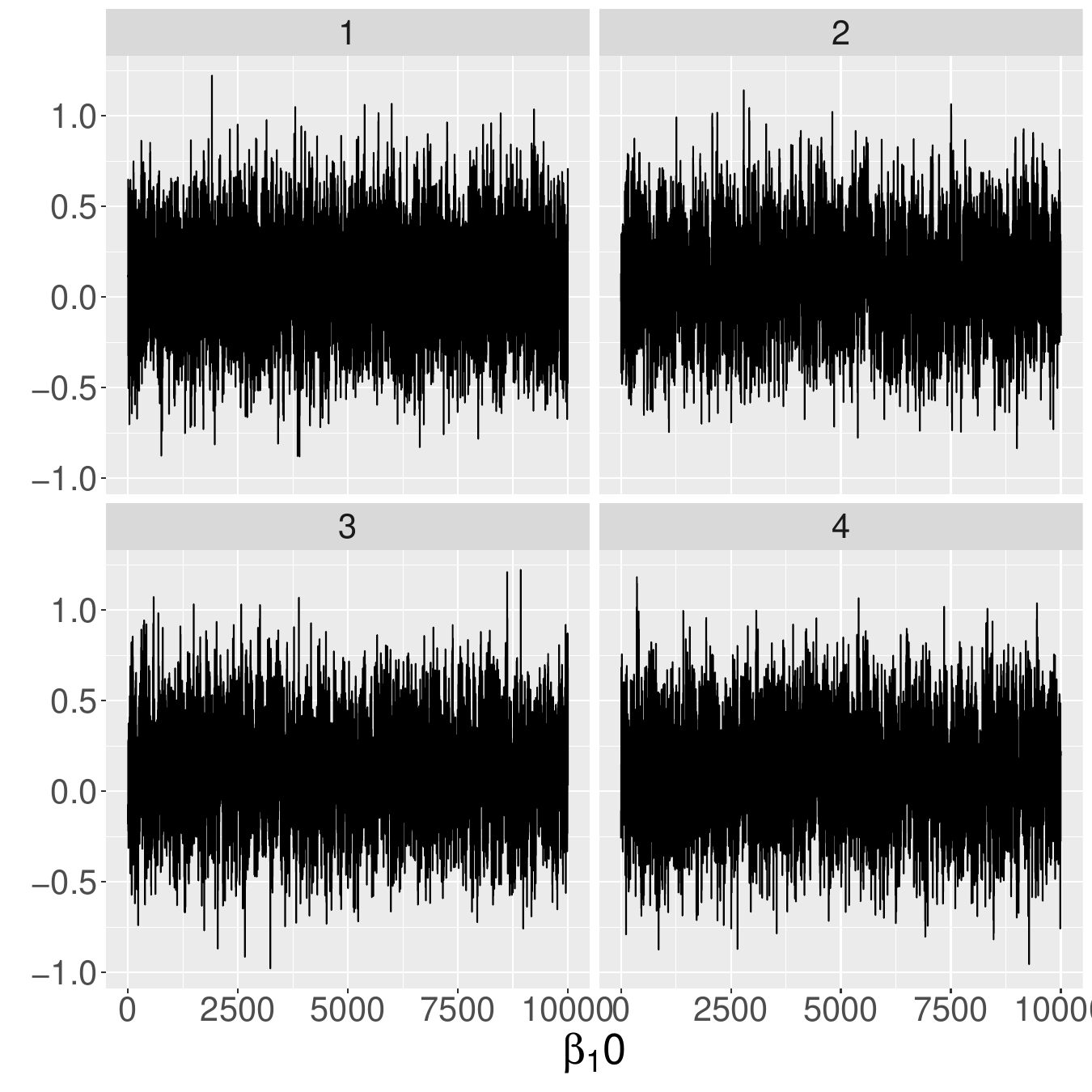}
        \caption{Individual 100}
    \end{subfigure}
    \begin{subfigure}[t]{0.3\textwidth}
        \includegraphics[width = \textwidth]{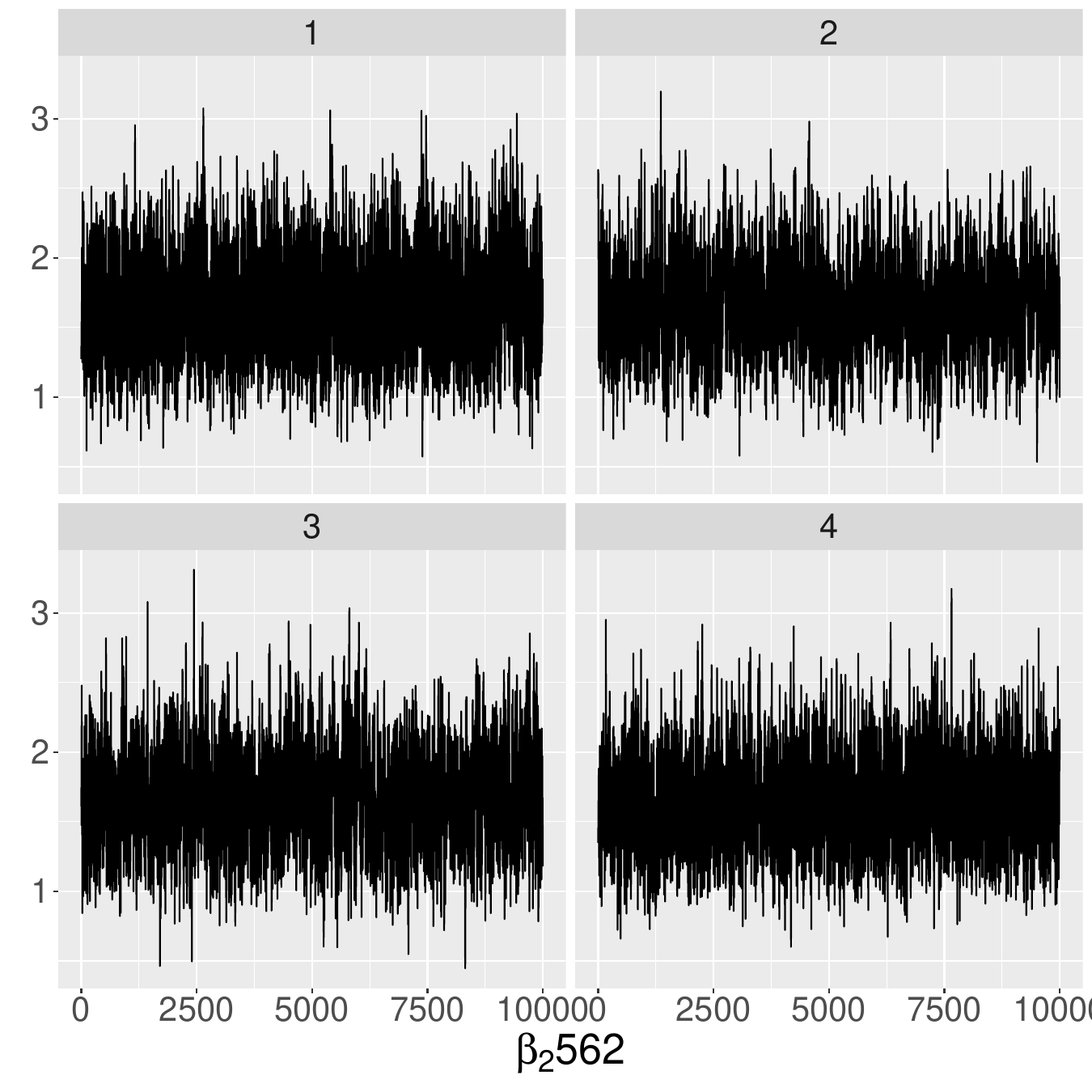}
        \caption{Individual 2562}
    \end{subfigure}\\
    \centering
    \begin{subfigure}[t]{0.3\textwidth}
        \includegraphics[width = \textwidth]{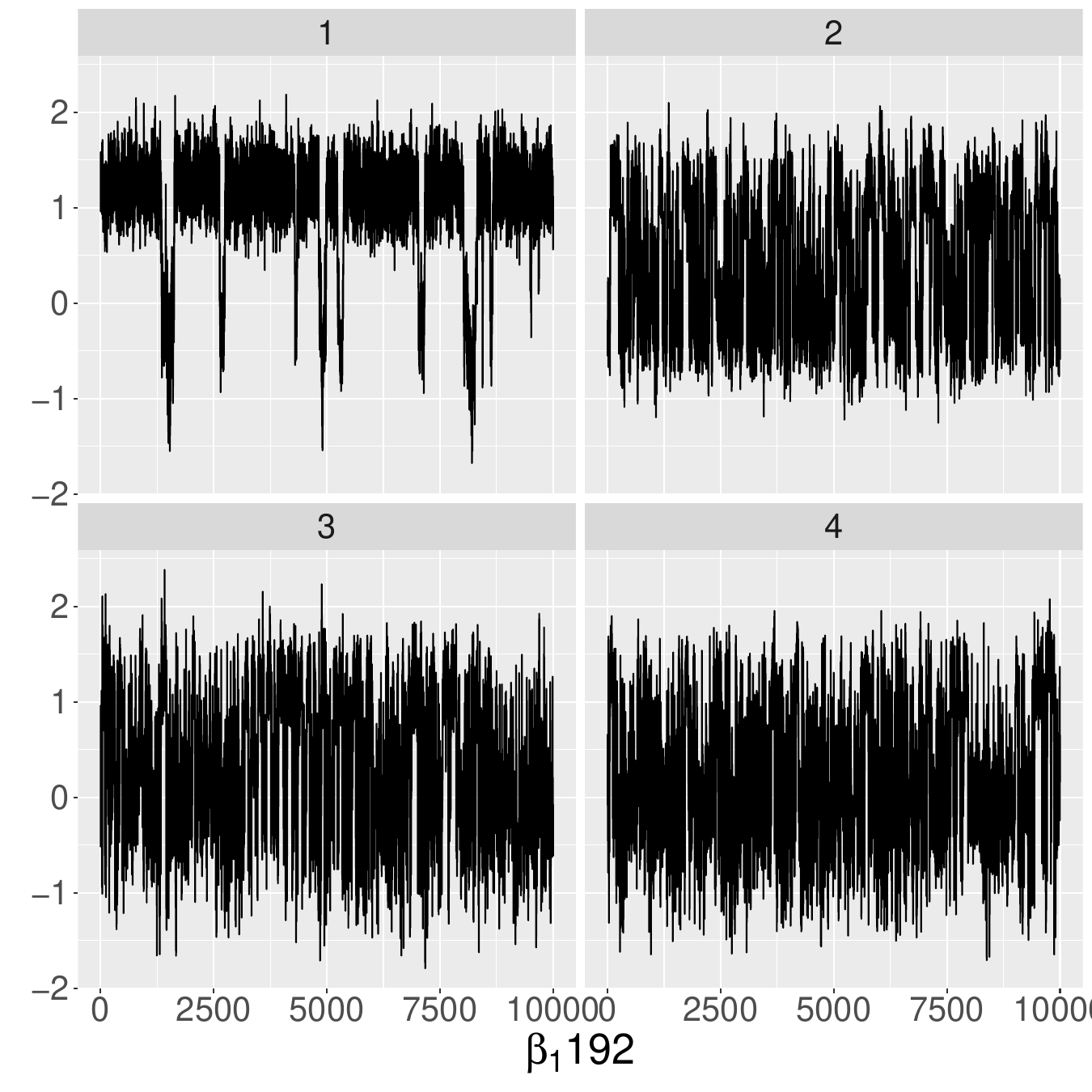}
        \caption{Individual 1192}
    \end{subfigure}
    \begin{subfigure}[t]{0.3\textwidth}
        \includegraphics[width = \textwidth]{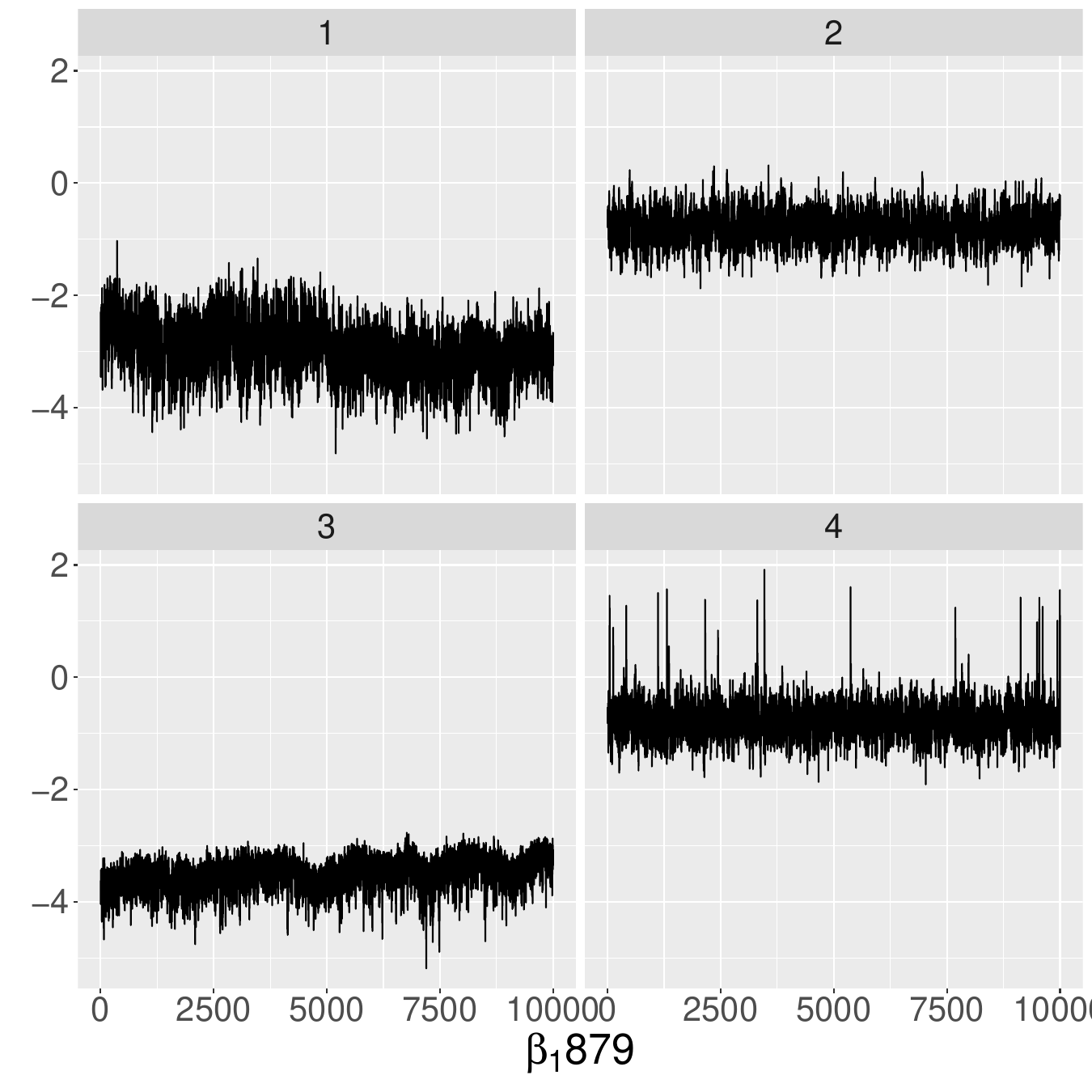}
        \caption{Individual 1879}
    \end{subfigure}
    \begin{subfigure}[t]{0.3\textwidth}
        \includegraphics[width = \textwidth]{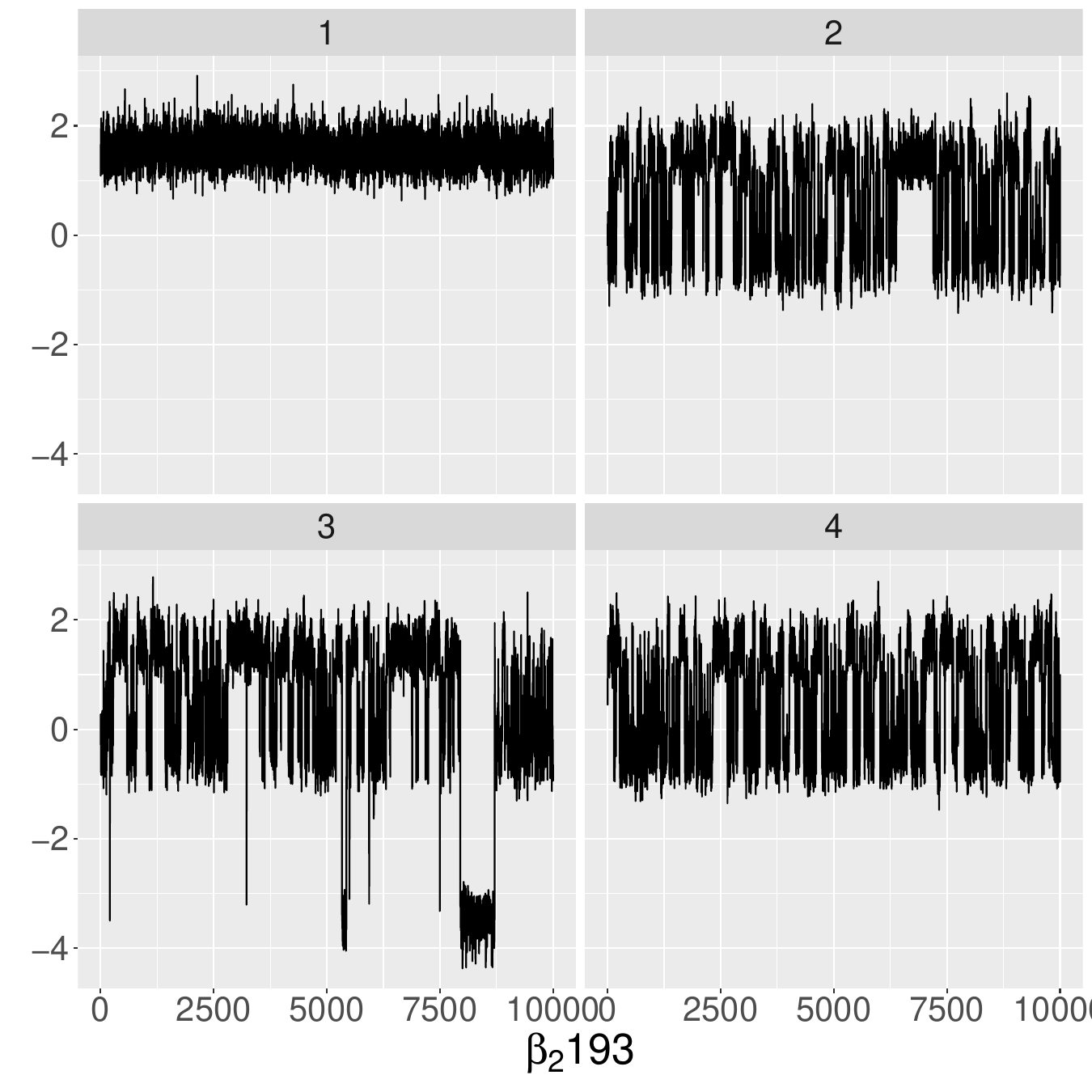}
        \caption{Individual 2193}
    \end{subfigure}
    \caption{Posterior trace plots of the latent trait $\beta_i$ for selected individuals.}
    \label{fig:betaposterior}
\end{figure}


\newpage

\section{Information function}
Our goal is to derive the information function for a single question $j$ and individual $i$ based on the individual's preference, $\beta_i$. In this case, the log likelihood function is defined to be:
\begin{align*}
    L(y_{i, j} \mid \beta_i, \bfalpha_j, \bfdelta_j) =  \sum_{k = 0}^{K_j} \mathbbm{I}(\zeta_j = k) \log(\theta_{i, j}(k)).
\end{align*}

As a result, we have that
\begin{align*}
    \frac{d}{d \beta_i} L(y_{i, j} \mid \beta_i, \bfalpha_j, \bfdelta_j) &= \sum_{k = 0}^{K_j} \mathbbm{I}(\zeta_j = k) \frac{1}{\theta_{i, j}(k)} \frac{d \theta_{i, j}(k)}{d \beta_i}
\end{align*}
and 
\begin{align*}
    \frac{d^2}{(d \beta_i)^2} L(y_{i, j} \mid \beta_i, \bfalpha_j, \bfdelta_j) &= \sum_{k = 0}^{K_j} \mathbbm{I}(\zeta_j = k) \left(-\frac{1}{\theta^2_{i, j}(k)} \left(\frac{d \theta_{i, j}(k)}{d \beta_i}\right)^2 + \frac{1}{\theta_{i, j}(k)} \frac{d^2 \theta_{i, j}(k)}{(d \beta_i)^2}\right).
\end{align*}

Because $E(\mathbbm{1}(\zeta_j = k)  \mid \beta_i) = \theta_{i, j}(k)$, then
\begin{align}
    E\left(\frac{d^2}{(d \beta_i)^2} L(y_{i, j} \mid \beta_i, \bfalpha_j, \bfdelta_j)\right) &= \sum_{k = 0}^{K_j} -\frac{1}{\theta_{i, j}(k)} \left(\frac{d \theta_{i, j}(k)}{d \beta_i}\right)^2 + \frac{d^2 \theta_{i, j}(k)}{(d \beta_i)^2}.
    \label{eq:info_fct_prob}
\end{align}

We can use the chain rule to explicitly compute the derivatives in \eqref{eq:info_fct_prob}. To do so, we need to use the rules of conditional probability and normal distribution to re-write $\phi_{2K_j}\left(
\v{z}^{-k} \mid \v{0}_{2K_j}, \bfSigma_j \right)$ for some $k \in -K_j, -K_j + 1, \ldots, K_j$, $k' \in -K_j, -K_j + 1, \ldots, K_j$, $k' \neq k$.
Here, the parameters are defined in the same manner as the parameters in the main manuscript right before Section 3.1. Then,
note that 
\begin{itemize}
    \item $[\bfSigma^{-1}_j]_{l, l'} = 1 - 1 / (2K_j + 1)$ when $l = l'$ and $[\bfSigma^{-1}_j]_{l, l'} = -1 / (2K_j + 1)$ when $l \neq l'$,
    \item $\mathbbm{1}_{2 K_j - 1} \bfSigma^{-1}_j = \frac{2}{2 K_j + 1} \mathbf{1}_{2 K_j}^T$, and
    \item $\mathbbm{1}_{2 K_j - 1} \bfSigma^{-1}_j \mathbf{1}_{2 K_j - 1} = \frac{2(2K_j - 1)}{2K_j + 1}$.
\end{itemize}
Here, we let $\bf{1}_{D}$ denote a vector of 1's of length $D \in \mathbbm{N}$. As a result,
\begin{multline*}
    \phi_{2K_j}\left(
    \v{z}^{-k} \mid \v{0}_{2K_j}, \bfSigma_j \right) =\\ 
    \phi_{1}\left(
    z^{-k}_{k'} \mid \frac{2}{2K_j + 1} \sum_{l = 1}^{2 K_j - 1} z^{-k, k'}_{l}, \frac{4}{2K_j + 1} \right)
    \phi_{2K_j - 1}\left(
    \v{z}^{-k, k'} \mid \v{0}_{2K_j - 1}, \bfSigma^{-k'}_j \right).    
\end{multline*}
Here, $\v{z}^{-k, k'} \in \mathbbm{R}^{2K_j - 1}$ denotes the same vector as $\v{z}^{-k}$ except with $z^{-k}_{k'}$ removed and $\bfSigma^{-k'}_j$ denotes the same matrix as $\bfSigma_j$ with the $j^{th}$ column and row taken out. Further, to make our notation more compact, let 
\[
\widetilde{\phi}_1(z^{-k}_{k'} \mid \v{z}^{-k, k'}, d) = \phi_{1}\left(
    z^{-k}_{k'} \mid \frac{2}{d + 1} \sum_{l = 1}^{d - 1} z^{-k, k'}_{l}, \frac{4}{d + 1} \right).
\]

This enables us to write the derivatives. For another simplification of notation, let
\begin{multline*}
\int_{-k, k'} =
     \int_{\eta_{i, -K_j} - \eta_{i, k}}^{\infty} \int_{\eta_{i, -K_j + 1} - \eta_{i, k}}^{\infty} \cdots \int_{\eta_{i, k - 1} - \eta_{i, k}}^{\infty} \int_{\eta_{i, k + 1} - \eta_{i, k}}^{\infty} \cdots \\
     \int_{\eta_{i, k' - 1} - \eta_{i, k}}^{\infty} \int_{\eta_{i, k' + 1} - \eta_{i, k}}^{\infty} \cdots \int_{\eta_{i, K_j} - \eta_{i, k}}^{\infty} 
\end{multline*}
Then, following our discussion above and by the Fundamental Theorem of Calculus, we have that
\begin{multline*}
    \frac{d}{d \beta_i} \nu_{i, j}(k) =
    \sum_{\substack{k' = -K_j,\\ k' \neq k}}^{K_j}  \int_{-k, k'} \phi_{2K_j - 1}\left(
    \v{z}^{-k, k'} \mid \v{0}_{2K_j - 1}, \bfSigma^{-k'}_j \right)\\ 
    -\widetilde{\phi}_1(\eta_{i, k'} - \eta_{i, k} \mid \v{z}^{-k, k'}, 2 K_j) (\alpha_{j,k'} - \alpha_{j, k}) d \v{z}.    
\end{multline*}

As a result, 
\begin{align}
    \frac{d \theta_{i, j}(k)}{d \beta_i} &= 
    \begin{cases}
      \frac{d}{d \beta_i} \nu_{i,j}(0)  & k=0 , \\
      \frac{d}{d \beta_i} \nu_{i,j}(-k) + \frac{d}{d \beta_i}\nu_{i,j}(k)  & k \ge 1,
    \end{cases}
    \label{eq:info_fct_prob_first_deriv}
\end{align}

Meanwhile, for the second derivative, we can follow the same logic and extend the notation used previously. Then,
\begin{align*}
    \frac{d^2}{d^2 \beta_i} & \nu_{i, j}(k) =\\
    &\sum_{\substack{k' = -K_j,\\ k' \neq k}}^{K_j} \sum_{\substack{k'' = -K_j,\\ k'' \neq k, k'}}^{K_j}  \int_{-k, k', k''} \phi_{2K_j - 2}\left(
    \v{z}^{-k, k', k''} \mid \v{0}_{2K_j - 2}, \bfSigma^{-k', k''}_j \right)\\ 
    & \qquad\qquad\qquad\qquad\qquad\qquad\quad  \widetilde{\phi}_1(\eta_{i, k''} - \eta_{i, k} \mid \v{z}^{-k, k', k''}, 2K_j - 1)\widetilde{\phi}_1(\eta_{i, k'} - \eta_{i, k} \mid \v{z}^{-k, k'}, 2 K_j)\\  
    & \qquad\qquad\qquad\qquad\qquad\qquad\qquad (\alpha_{j,k''} - \alpha_{j, k})(\alpha_{j,k'} - \alpha_{j, k}) d \v{z} +\\
    & \qquad\qquad \int_{-k, k'} \phi_{2K_j - 1}\left(
    \v{z}^{-k, k'} \mid \v{0}_{2K_j - 1}, \bfSigma^{-k'}_j \right) \widetilde{\phi}_1(\eta_{i, k'} - \eta_{i, k} \mid \v{z}^{-k, k'}, 2 K_j)\\ 
    & \qquad\qquad\qquad\qquad \frac{\eta_{i, k'} - \eta_{i, k} - \sum_{l = 1}^{2 K_j - 1} z^{-k, k'}_{l}}{2 - \frac{2 K_j}{2K_j + 1}} (\alpha_{j,k'} - \alpha_{j, k})^2 d \v{z}.    
\end{align*}

Hence, this gives us the second derivative, i.e.
\begin{align}
    \frac{d^2}{d^2 \beta_i} \nu_{i, j}(k) &= 
    \begin{cases}
      \frac{d^2}{d^2 \beta_i} \nu_{i,j}(0)  & k=0 , \\
      \frac{d^2}{d^2 \beta_i} \nu_{i,j}(-k) + \frac{d}{d \beta_i}\nu_{i,j}(k)  & k \ge 1,
    \end{cases}
    \label{eq:info_fct_prob_second_deriv}
\end{align}

As a result, if we substitute \eqref{eq:info_fct_prob_first_deriv} and \eqref{eq:info_fct_prob_second_deriv} into \eqref{eq:info_fct_prob} and ensure that \eqref{eq:info_fct_prob} is positive, we get the information function.

\end{document}